\documentclass[twocolumn, times, tighten]{aastex62}
\usepackage{textcomp}
\usepackage{amsmath}
\usepackage{color}
\usepackage{amssymb}
\usepackage{hyperref}
\usepackage[caption=false]{subfig}

\newcommand{\mum}{\ifmmode{\rm \mu m}\else{$\mu$m\ }\fi}

\shorttitle{The intrinsic AGN IR SED}

\begin{document}
\revised{\bf Draft: \today}

\title{AGNs are not that cool: \\
    revisiting the intrinsic AGN far-infrared spectral energy distribution}

\author{Jun Xu}
\affiliation{CAS Key Laboratory for Research in Galaxies and Cosmology,
Department of Astronomy, University of Science and Technology of China,
Hefei 230026, China; ericsun@ustc.edu.cn; xuey@ustc.edu.cn}
\affiliation{School of Astronomy and Space Science, University of Science
and Technology of China, Hefei 230026, China}

\author[0000-0002-0771-2153]{Mouyuan Sun}
\affiliation{CAS Key Laboratory for Research in Galaxies and Cosmology,
Department of Astronomy, University of Science and Technology of China,
Hefei 230026, China; ericsun@ustc.edu.cn; xuey@ustc.edu.cn}
\affiliation{School of Astronomy and Space Science, University of Science
and Technology of China, Hefei 230026, China}
\affiliation{Department of Astronomy, Xiamen University, Xiamen, Fujian 
361005, China}

\author[0000-0002-1935-8104]{Yongquan Xue}
\affiliation{CAS Key Laboratory for Research in Galaxies and Cosmology,
Department of Astronomy, University of Science and Technology of China,
Hefei 230026, China; ericsun@ustc.edu.cn; xuey@ustc.edu.cn}
\affiliation{School of Astronomy and Space Science, University of Science
and Technology of China, Hefei 230026, China}

\begin{abstract}
We investigate the intrinsic spectral energy distribution (SED) of active galactic
nuclei (AGNs) at infrared (IR) bands with 42 $z < 0.5$ optically luminous
Palomar Green survey quasars through SED decomposition. We decompose
the SEDs of the 42 quasars by combining an AGN IR template library
\citep{Siebenmorgen2015} that covers a wide range of the AGN parameter space
with three commonly used galaxy template libraries. We determine 
the median AGN SED from the best-fitting results. The far-IR (FIR) contribution 
of our median AGN SED is significantly smaller than that of 
\cite{Symeonidis2016}, but roughly consistent with that of \cite{Lyu2017}. The 
AGN IR SED becomes cooler with increasing bolometric luminosity, 
which might be due to that more luminous AGNs might have 
stronger radiative feedback to change torus structures and/or their tori 
might have higher metallicities. 
Our conclusions do not depend on the choice of galaxy template 
libraries. However, since the predicted polycyclic aromatic 
hydrocarbon (PAH) emission line flux is galaxy template-dependent, 
cautions should be taken on deriving galaxy FIR contribution from PAH fluxes. 
\end{abstract}


\section{Introduction}
It has been a common view that supermassive black holes (SMBHs) lie at the 
centers of typical massive galaxies. SMBHs swallow nearby gas through 
accretion disks (thereby being active galactic nuclei ---AGNs) and their hosts 
form stars from cold gas at large scales. In this scenario, SMBHs 
become more massive and the accreted material radiates across a wide range 
of the electromagnetic spectrum. This is known as the quasar phase. 
According to the unified model \citep[e.g.,][]{Antonucci1993, 
Urry1995}, AGNs are believed to be surrounded by dusty tori 
\citep[e.g.,][]{Tristram2007}. The dusty torus can absorb AGN UV/optical emission 
and re-radiate in the near-infrared, middle-infrared (MIR) and far-infrared (FIR) 
bands. The IR emission from the heated dusty torus usually peaks at MIR 
bands \citep[e.g.,][]{Antonucci1993} and turns over at $\sim 20$--$70\ 
\mathrm{\mu m}$ \citep[e.g.,][; \citealt{Mullaney2011}, hereafter M11]{Netzer2007}.

AGNs can have significant influence on their hosts \citep[i.e., AGN 
feedback; see e.g.,][]{Silk1998, King2003, Fabian2012} by strong multi-band 
radiation, multi-scale and multi-phase winds \citep[e.g.,][]{Blandford1982, 
Murray1995, Reynolds1997, Proga2004, Trump2006, Richards2011, 
Yuan2012, Cao2013, Tombesi2013, Filiz2014, Grier2015, Gu2015, Mou2017, 
Sun2018, Sun2019, He2019} and/or relativistic jets \citep[e.g.,][]{McNamara2000}. 
They can heat the interstellar medium (ISM) and/or eject the ISM and thereby reduce 
or even quench star formation in the host galaxies. Such a process might 
be able to explain the well-established $M_{\rm{BH}}-\sigma$ 
\citep[e.g.,][]{Ferrarese2000, Gebhardt2000, Tremaine2002, Kormendy2013} and 
$M_{\rm{BH}}-M_{\rm{bulge}}$ \citep[e.g.,][]{Kormendy1992, Magorrian1998, 
Haring2004, Gultekin2009} scaling relations. To reveal the physical nature 
of the relations between SMBHs and their hosts, we should accurately measure 
the properties of AGNs (e.g., $M_{\rm{BH}}$, accretion rate) and their host 
galaxies (e.g., star-formation rate, stellar mass) across cosmic history 
\citep[e.g.,][]{Sun2015}.

Star-formation rates (SFRs) are often measured \citep[for a review, 
see][]{Kennicutt1998} by the total luminosity of ultraviolet (UV) emission 
of a galaxy or some recombination lines (e.g., H$\alpha$). However, such 
estimators can be strongly contaminated by AGN emission. Stellar UV emission 
might also be absorbed by dusty clouds in galaxies. Unlike AGN emission, 
the stellar emission is usually much fainter. Thus, the average dust temperature 
is much cooler \citep[e.g.,][]{Elvis1994, Richards2006, Netzer2007,
Mullaney2011} and the re-radiated IR emission peaks at FIR bands 
\citep[e.g.,][]{Dale2002}. Consequently, it has been suggested that FIR bands 
provide a clean window for measuring star-formation activities of host 
galaxies \citep[e.g.,][]{Page2012, Harrison2012}.

However, it was claimed \cite[][hereafter S16]{Symeonidis2016} that the AGN 
emission in FIR bands is not negligible; without properly removing the AGN 
contamination, SFRs of many AGN hosts are significantly overestimated. 
S16 used a sample of 47 broad-line, luminous (the luminosity at $5100\ \rm{\AA}$ 
$L_{5100}>10^{43.5}\; \rm{erg\;s^{-1}}$), $z<0.18$, radio-quiet quasars from the
Palomar Green survey \citep[hereafter PG quasars;][]{Schmidt1983}. By using
the archival data in the $0.4$--$500$ $\rm{\mu m}$ range, S16 constructed the 
IR spectral energy distributions (SEDs). They derived the stellar contribution 
by matching the strength of $11.3\ \rm{\mu m}$ polycyclic aromatic hydrocarbon (PAH)
feature with the galaxy template library of \citet[][hereafter DH02]{Dale2002}, i.e., 
selecting the galaxy template with the $11.3\ \rm{\mu m}$ PAH strength that is closest 
to the observed value. S16 then subtracted the average stellar 
contribution from the average PG quasar SED, resulting in the 
intrinsic AGN IR SED. Their intrinsic AGN IR SED 
is more luminous at FIR bands (by $\sim 0.5$ dex; i.e., the average dust temperature is much 
cooler) than those of previous works \citep[e.g.][M11]{Netzer2007} for fixed $20\ 
\rm{\mu m}$ emission. Meanwhile, S16 found that AGNs with different 11.3 \mum PAH 
luminosities tend to share the same FIR profile. This result is also incompatible with 
that of M11 who found that luminous AGNs tend to have a lower ratio of FIR emission 
to the total IR emission (see M11 Figure 6). S16 argued that the cool FIR emission 
could be a result of AGN UV/optical emission heating up the galactic-scale dust. If 
correct, the S16 results indicate that our understanding of star formation 
in AGN hosts would be substantially modified. 

The DH02 galaxy template library is constructed as follows. \cite{Dale2001} 
developed a new phenomenological model characterized by a 
single parameter $f_\nu^{60\;\mathrm{\mum}}/f_\nu^{100\;\mathrm{\mum}}$ for normal 
star-forming galaxies. The models were constrained by $Infrared\;Astronomical\;Satellite$ 
(IRAS) and $Infrared\;Space\;Observatory$ (ISO) broadband photometric data of 69 normal 
galaxies with different IR luminosities. They replaced the \cite{DBP90} PAH emission 
profiles with actual data from ISO. Their spectra extending up to $11\ \mum$ show an 
invariant shape regardless of their infrared-to-blue ratios. This indicates 
that there is no prominent $9.7$ $\rm{\mum}$ silicate absorption in their SEDs. 
Later, DH02 modified the FIR/submillimeter dust emissivity in the models to 
consider different radiation-field intensities and to match the long-wavelength (i.e., 
$\sim 100$--$800\ \rm{\mu m}$) data; then, they derived a commonly used 
galaxy template library. After that, \citet[][hereafter DH14]{Dale2014} updated the 
star-forming galaxy template library by considering the new $Spitzer$ 
high-quality Infrared Spectrograph (IRS\footnote{For more details about IRS, please refer 
to \url{http://irsa.ipac.caltech.edu/data/SPITZER/docs/irs/}.}) spectra. Their new 
templates also contain an AGN component whose 
relative strength can be varied. In this work, we set the templates to zero
AGN contribution. Meanwhile, \citet[][hereafter R09]{Rieke2009} assembled a galaxy
template library by using a sample of eleven local luminous IR galaxies (LIRGs) and
ultra-luminous IR galaxies (ULIRGs). They constructed the near- to mid-IR profile (including
the PAH and silicate features) by taking advantage of $Spizter$ IRS spectra \citep{Houck2004} 
and ISO \citep{Rigopoulou1999} spectra. At FIR bands, a modified blackbody model was
fitted to the FIR and submillimeter photometry of each galaxy. In contrast to \cite{Dale2002}
and \cite{Dale2014}, the $9.7$ $\rm{\mum}$ silicate absorption feature is included 
in the R09 library. 

Recently, \cite{Lani2017} and \citet[][hereafter L17]{Lyu2017} checked the result of S16 
by following the S16 PAH strength based method. However, they failed to obtain the AGN 
FIR SED of S16. Unlike S16, \cite{Lani2017} (who also used the DH02 templates) normalized 
the individual galaxy-emission-subtracted SEDs at $20$ \mum before deriving 
the average AGN SED (see also Section~\ref{sect:sed}). L17 used the R09 templates instead, 
and took advantage of \cite{Elvis1994} sample, then used the relation between the 11.3 
$\mu$m PAH equivalent width and the 25 $\mu$m to 60 $\mu$m or 24 \mum to 70 \mum flux 
ratio to estimate FIR contribution of star formation in \cite{Elvis1994}.
L17 argued that the stellar contribution would be biased to low values if the DH02 templates 
were adopted for PG quasars. This is because, as pointed out by L17, the $11.3$ \mum 
PAH line may overlap with the $9.7$ \mum silicate absorption feature. If the $9.7$ \mum \ 
silicate absorption is strong and ignored, the measured $11.3$ \mum \ PAH flux will be smaller 
than the true flux. As a result, the stellar emission inferred from the $11.3$ \mum \ PAH 
flux will also be underestimated. In addition, M11 constructed the AGN IR SED in an empirical 
way: first, they assumed that the AGN IR SED can be described by a modified blackbody 
function; second, they adopted this modified blackbody function (with unknown parameters, 
e.g., the blackbody temperature) and five starburst galaxy templates to simultaneously 
fit the IRS spectra and \textit{IRAS} data of $11$ \textit{Swift}-BAT AGNs. The mean 
of the best-fitting AGN SEDs is also inconsistent with that of S16 (see 
Section~\ref{sect:result}).

In this work, we aim to check the result of S16 via a two-component (i.e., a galaxy 
component plus an AGN component) SED decomposition method. We use the AGN templates 
of \citet{Siebenmorgen2015}, because, for a fixed MIR luminosity, their FIR 
luminosities can reproduce the popular AGN IR SEDs (see Figure~\ref{fig:templates}).

This paper is laid out as follows. In Section~\ref{sect:data}, we describe our sample 
and data.In Section~\ref{sect:method}, we introduce our decomposition method. In
Section~\ref{sect:result}, we show our new AGN FIR SEDs and discuss our results. 
Conclusions are drawn in Section~\ref{sect:conclusions}. Throughout
this work, we adopt a flat $\Lambda$CDM cosmology of $\mathrm{H}_0=70\;
\mathrm{km\,s^{-1}\,Mpc^{-1}}$ and $\Omega_M=0.3$ \citep{Peebles2003}.

\section{Sample and Data}
\label{sect:data}
\begin{figure}
\epsscale{1.1}
\plotone{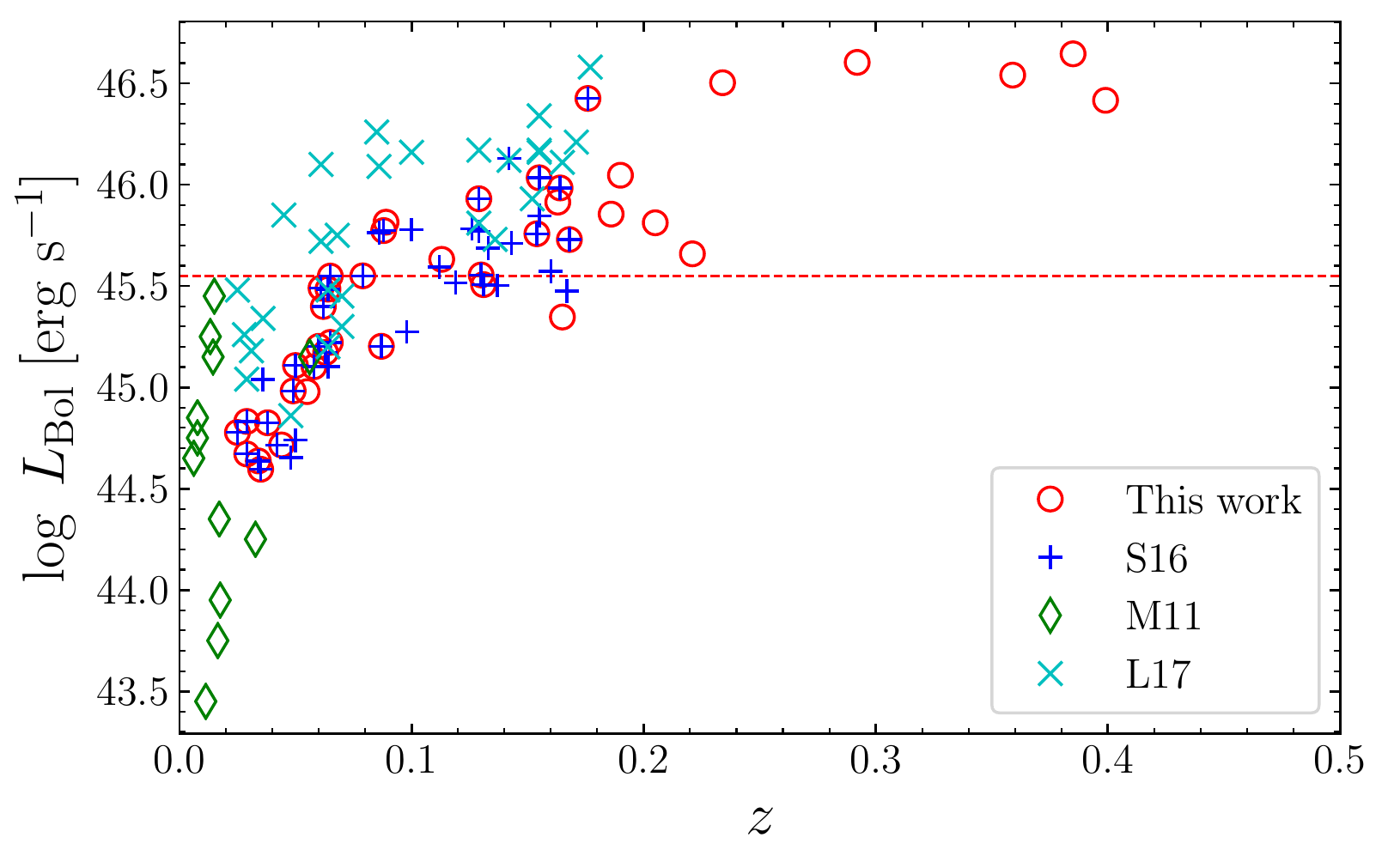}
\caption{The bolometric luminosity as a function of redshift for our selected
PG quasars (42 sources; red open circles), the red dashed line represents the 
median bolometric luminosity of our sample. For comparison, we also include the 
AGN samples of S16 (47 sources; blue pluses), M11 (12 sources; green diamonds), 
and L17 (29 sources; cyan crosses).}
\label{fig:L_z}
\end{figure}

Our goal is to verify the result of S16. Therefore, we also adopt the PG quasar 
sample. Our
parent sample includes all 87 objects of the PG quasar sample \citep[e.g.,][]{Schmidt1983,
Boroson1992} at redshift $z < 0.5$ (see Figure~\ref{fig:L_z}). We collect the multi-band
photometric data as follows: Palomar $B$-band photometry \citep[e.g.,][]{Schmidt1983,
Shi2014}, SDSS magnitudes from the SDSS Photometric Catalog \citep[Release 9;][]{Ahn2012},
WISE magnitudes from the AllWISE Source Catalog \citep{Wright2010}, Spitzer/MIPS data
from \cite{Shi2014}, and the $Herschel$/SPIRE \citep{Griffin2010} data from \cite{Petric2015}.
Unlike S16, we exclude the photometric data from 2MASS, ISO and \textit{AKARI}. 
These data are out of
date compared with $Spitzer$ and $Herschel$ data. We also collect the high-quality
\textit{Spizter} IRS spectra for the 87 quasars from \cite{Shi2014}. The spectra are 
re-binned into 2$\mum$ intervals, starting from $5.7\ \mum$. The 
re-binned\footnote{By re-binning the spectra, the effects 
of emission/absorption lines on the subsequent SED decomposition can be diluted. } 
data can provide observed-frame $6$-$30$ \mum\ SEDs and are used in the subsequent 
SED decomposition. 

\begin{table}
\begin{center}
\caption{Adopted parameter space of the \cite{Siebenmorgen2015} AGN template library
\label{tbl-1}}
\begin{tabular}{c|c}
\tableline
\tableline
$\theta$ (degree) &  19, 33, 43, 52, 60, 67\\
\tableline
$R_s$ ($10^{15}\rm{cm}$) &   300, 514, 772, 1000, 1545\\
\tableline
$V_c$ (\%) & 1.5, 7.7, 38.5, 77.7\\
\tableline
$A_c$ & 0, 4.5, 13.5, 45\\
\tableline
$A_d$ & 0, 30, 100, 300, 1000\\
\tableline
\end{tabular}
\tablecomments{$\theta$, $V_c$, $A_c$, and $A_d$ represent the viewing angle, the cloud 
volume filling factor, the optical depth of an individual cloud, and the optical depth of the disk 
mid-plane, respectively. $R_s$ corresponds to the inner radius of the dusty torus 
if an AGN luminosity with $L_{\mathrm{bol}}=10^{11}\ L_{\odot}$, where $L_{\odot}$ is the solar 
luminosity. That is, larger $R_s$ indicates colder dust sublimation temperature (see texts).}
\end{center}
\end{table}

We removed $16$ radio-loud \citep[with radio loudness $\mathrm{R}>10$; 
see][]{Kellermann1989} sources to avoid the potential contamination from jets. 
We also dropped two sources without available radio loudness estimates. 
For the remaining 69 radio-quiet PG quasars, we rejected 23 sources without 
observations at rest-frame wavelength $\lambda_{\rm{rest}}>200\
\rm{\mu m}$. Thus, the sample for our subsequent analysis consists of 46 quasars.
Following S16, we interpolate between the SDSS $ugriz$ bands or 
between the Palomar B-band and 2MASS J-band (in the cases of no SDSS counterparts) 
to derive the rest-frame $5100\ \mathrm{\AA}$ luminosity (i.e., $\lambda L_{\lambda}$ at 
5100 \AA, hereafter $L_{5100}$). A comparison between the optical-derived 
and IR-derived rest-frame $5100\ \mathrm{\AA}$ luminosities is discussed in 
the appendix Section~\ref{sect:energy}. 

The distribution of our sample in the $L_{\rm{bol}}$--$z$ plane is presented 
in Figure~\ref{fig:L_z}. $L_{\rm{bol}}$ is estimated from $L_{5100}$ with a bolometric 
factor of $10$ \citep{Richard2006}. For comparison purpose, we also 
show the distributions of samples used 
in previous works in Figure~\ref{fig:L_z}. For each source in the S16 sample, 
$L_{\rm{bol}}$ is also estimated from $L_{5100}$. The AGN IR SED of L17 is 
based on the AGN sample of \cite{Elvis1994}; for each of these AGNs, \cite{Elvis1994} integrated 
the observed SEDs to derive $L_{\rm{bol}}$ (see their section 6.1 for details). For AGNs 
in the M11 sample, we use the hard X-ray luminosity 
$L_{\rm 2-10\ keV}$ and the bolometric correction of \cite{Lusso2012} to estimate 
$L_{\rm{bol}}$. The M11 AGNs are less luminous than those of our sample, S16 and L17.

\section{Spectral Energy Distribution decomposition}
\label{sect:method}
At FIR bands, the galaxy contribution is significant. To isolate the AGN component, we use
a two-component SED decomposition. That is, the observed fluxes are
\begin{equation}
\label{eq:f1}
f_{\mathrm{obs}} = c_1 f_{\mathrm{AGN}} + c_2 f_{\mathrm{Gal}},
\end{equation}
where $f_{\mathrm{AGN}}$ and $f_{\mathrm{Gal}}$ are the AGN and galaxy fluxes, respectively.
The remaining two free parameters are the normalizations of the two templates 
(i.e., $c_1$ and $c_2$).\footnote{With the inclusion of the two free parameters in the 
galaxy modified blackbody function (i.e., the temperature and 
power index of frequency) and the five free parameters in \cite{Siebenmorgen2015} AGN template 
library, the total number of free parameters in our SED fitting is nine.} 
It is worth noting that, for each source, the best-fitting galaxy component cannot 
exceed the observed SED; however, this is not the case in S16. 
The disadvantage of this method is that we need to use the shapes of known AGN SEDs 
as a prior. Therefore, we wish to use an AGN template library that covers a wide range of the AGN 
parameter space and also contains commonly used AGN IR SEDs \citep[e.g., the SEDs of][]{Mullaney2011, 
Symeonidis2016, Lyu2017}. 

\begin{figure*}
\epsscale{1.1}
\plotone{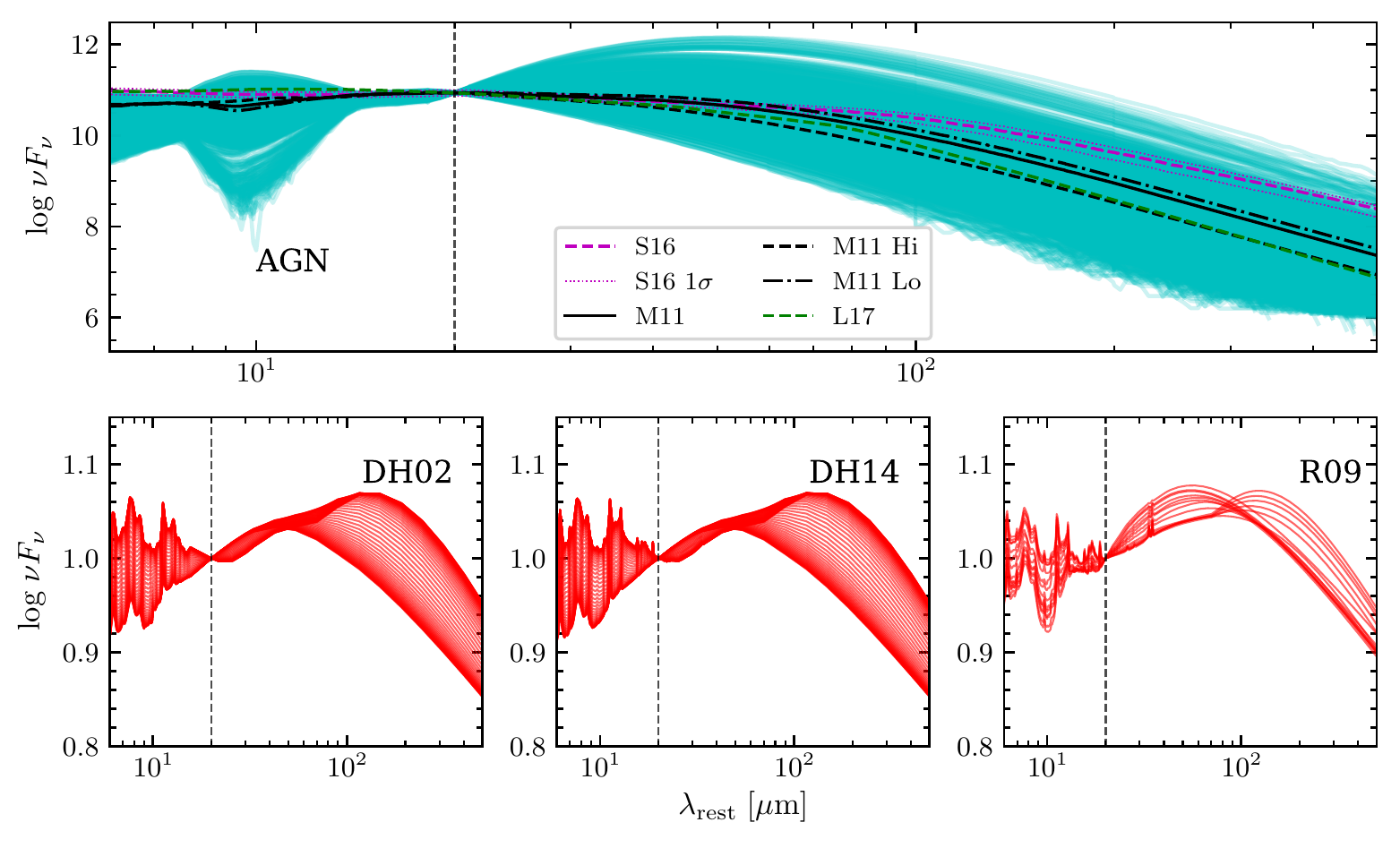}
\caption{The templates adopted in our SED decomposition. Upper panel: The \cite{Siebenmorgen2015}
AGN SED templates (thin cyan curves) and three previous AGN median SEDs (see the legend; 
the 1 $\sigma$ errors of the S16 SED, mean SEDs of the high-luminosity and low-luminosity 
subsamples of M11 are also included). Lower panels: three galaxy template libraries from DH02, 
DH14, and R09, respectively. All the templates are normalized at 20 $\mathrm{\mu}$m (i.e., the 
vertical dashed lines).}
\label{fig:templates}
\end{figure*}

Our AGN templates are selected from \cite{Siebenmorgen2015}.\footnote{For more details about
the \cite{Siebenmorgen2015} SED library of AGN torus models, please refer to
\url{http://www.eso.org/~rsiebenm/agn_models/index.html}.} They derived this template library 
by assuming that the dust clouds near an AGN are distributed in a torus-like geometry, which 
may be described by a clumpy medium, a homogeneous disk, or a combination of the two. They
considered the AGN dust structure to be approximated by an isothermal disk that is embedded in
a clumpy medium. The parameters of this library are $\theta$, $V_c$, $A_c$, and $A_d$
(see Table~\ref{tbl-1} for our adopted parameter space), which represent the viewing angle (in
degrees) measured from the pole ($z$-axis), the cloud volume filling factor, the $V$-band optical 
depth of an individual cloud, and the $V$-band optical depth of the disk mid-plane, 
respectively. We do not consider templates with the viewing angle $\theta = 73,\ 80,\ 86$ since 
our sources are type-1 AGNs. The remaining free parameter is $R_s$, which is the inner 
radius of the dusty torus (in units of $10^{15}\;\rm{cm}$) for an AGN with fixed bolometric luminosity 
\citep[i.e., $10^{11} L_{\odot}$; see][]{Siebenmorgen2015}; that is, $R_s$ indicates the 
dust sublimation temperature (for more details, see Section 4.2). 
The templates we select can encompass commonly-used AGN SEDs (see the upper panel of
Figure~\ref{fig:templates}).

\begin{figure*}
\epsscale{1.1}
\includegraphics[width=0.5\linewidth]{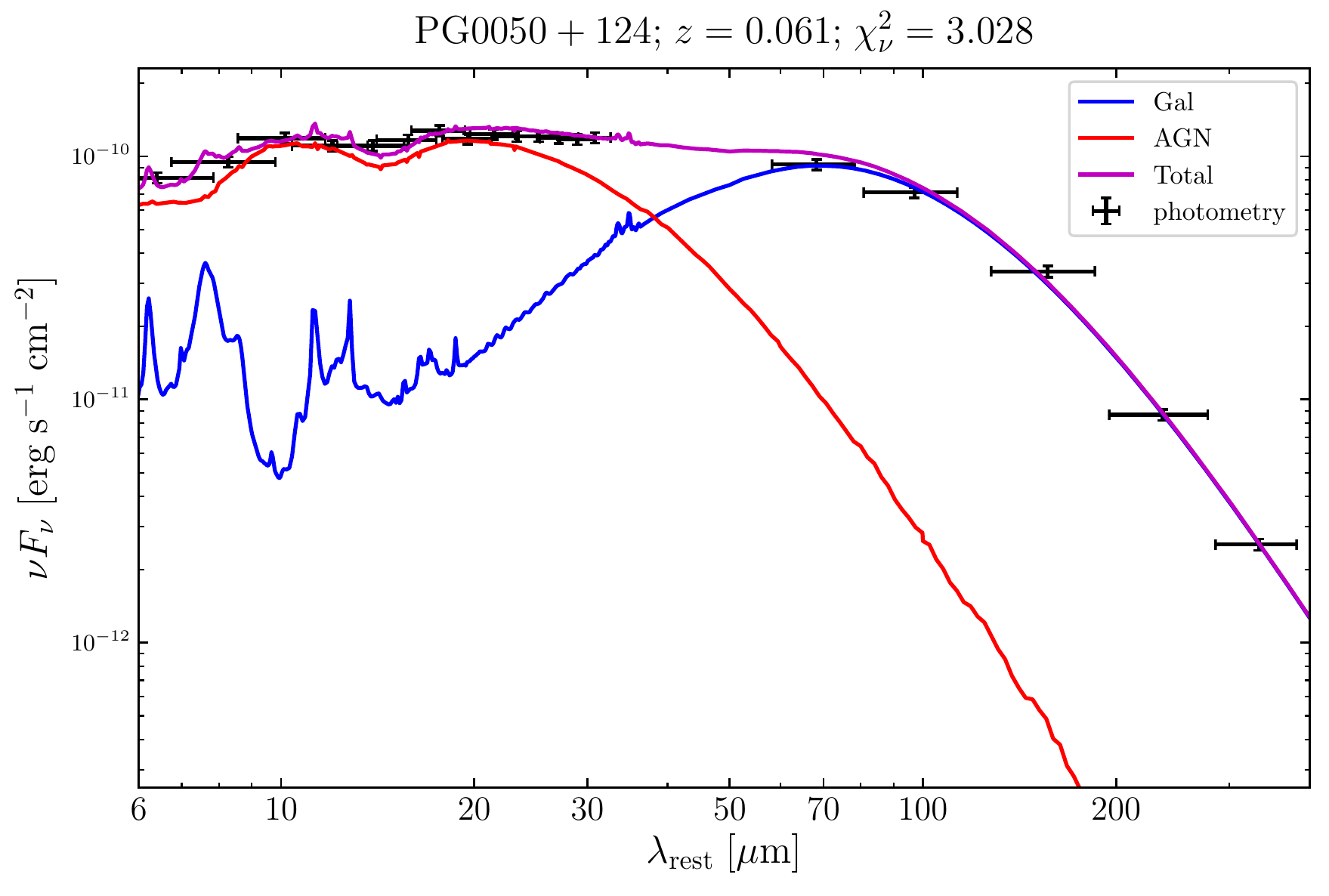}
\includegraphics[width=0.5\linewidth]{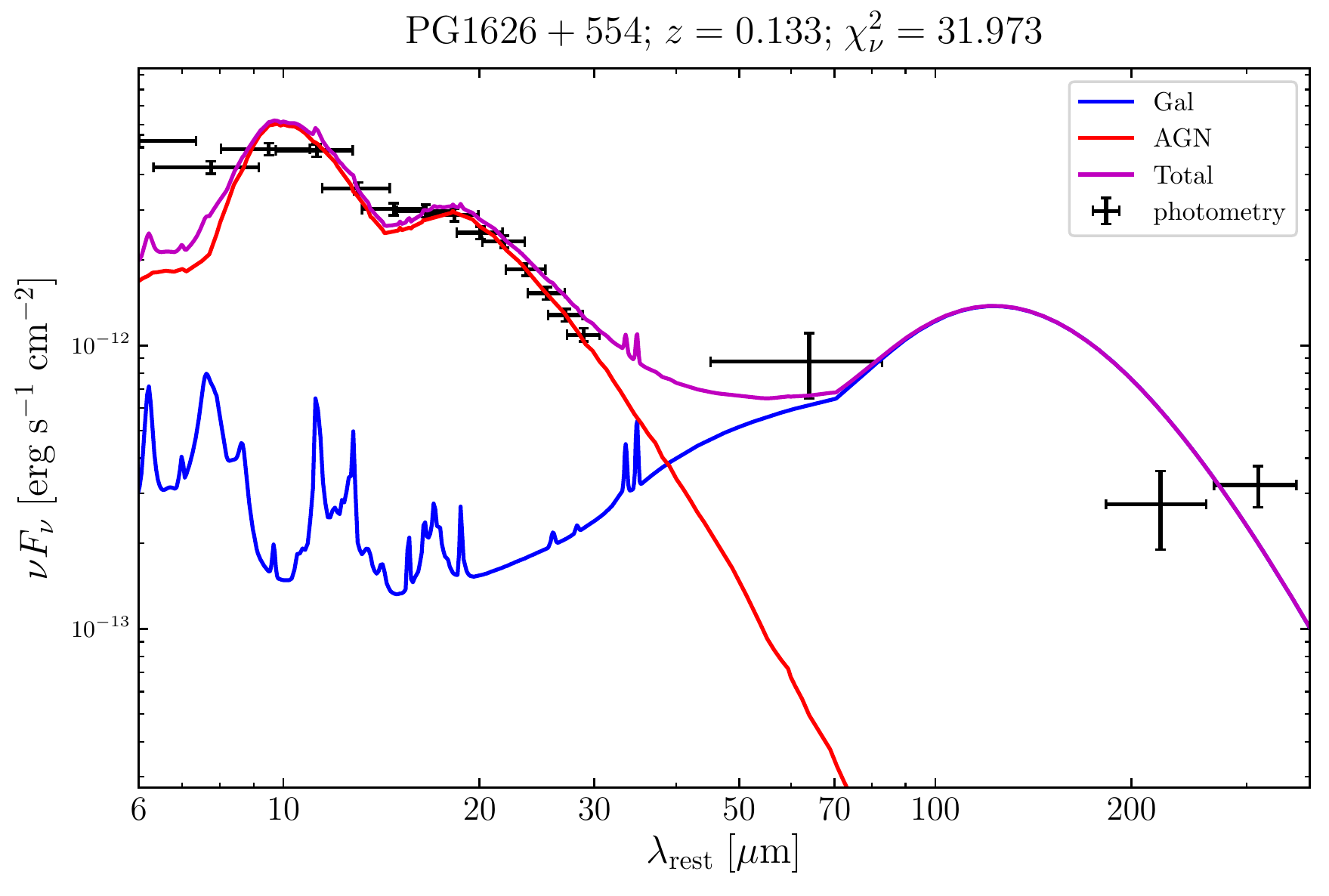}
\caption{An illustration of our SED decomposition results using the R09 templates. \texttt{PG0050+124}
is an example of reliably fitted sources. The profile matches well to the observed SED.
\texttt{PG1626+554} is an example of unreliably fitted sources, whose best-fit profile does not match
the low-quality SED data points. Meanwhile, its observations have limited coverage at $\lambda 
> 200\ \mum$. The complete figure set (42 images) is available in the online 
journal. }
\label{fig:decomposition}
\end{figure*}

To test the likely dependency of our results on the choice of galaxy templates, here we consider 
three galaxy libraries: DH02, DH14, and R09 (see the lower panels of Figure~\ref{fig:templates}).
We use \texttt{SEABASs} \citep{Rovilos2014},\footnote{For more details about \texttt{SEABASs},
please refer to \url{http://xraygroup.astro.noa.gr/SEABASs}.} a bayesian SED-decomposition code,
to fit Eq.~\ref{eq:f1} to the multi-band data. For each source, this code calculates the 
corresponding ``local'' maximum likelihoods of all possible combinations of AGN and galaxy templates 
by varying the normalization factors $c_1$ and $c_2$ (via a Monte Carlo Markov Chain sampling 
algorithm). As a second step, the ``local'' maximum likelihoods of all possible combinations of 
templates are compared to determine the ``global'' maximum value. The template combination 
(and the best-fitting $c_1$ and $c_2$) that gives the ``global'' maximum likelihood is selected 
as the best-fitting one. During our SED decomposition, we do not consider any priors. For each source, 
we record the reduced chi-squared $\chi^2_\nu$ of the best-fitting result (see Table~\ref{tbl-3}), and 
pure AGN model fitted result as a comparison.
\begin{table}
\begin{center}
\caption{Reduced $\chi^2$ of our best fitting results.}
\label{tbl-3}
\begin{tabular}{|c|c|c|}
\hline
    Name       & $\chi^2_\nu$ & $\chi^2_\nu$\\
    ---          & (AGN+Gal)    & (AGN only)\\
    \hline
    PG0003+199 & 38.37 & 2363.87 \\
    PG0043+039 & 5.41 & 446.13\\
    PG0050+124 & 3.03 & 2273.21\\
    PG0052+251 & 14.62 & 511.47\\
    PG0157+001 & 4.58 & 2728.24\\
    PG0838+770 & 4.55 & 650.90\\
    PG2130+099 & 4.29 & 846.77\\
    \vdots.  & \vdots &\\
    PG2214+139 & 18.36 & 9694.58\\
\hline
\end{tabular}
\tablecomments{The full table can be seen in the online version.}
\end{center}
\end{table}

Two examples of our SED decomposition results are presented in Figure~\ref{fig:decomposition}.  
%
\figsetstart
\figsetnum{1}
\figsettitle{displays the two-component SED decomposition results for all our $42$ sources.}
\figsetgrpstart
\figsetgrpnote{The SED decomposition results of our $42$ sources using the R09
    templates. }
\figsetgrpend
\figsetend
For comparison, the best-fitting results with the AGN templates alone (i.e., by fixing $c_2$ in 
Eq.~\ref{eq:f1} to be zero) are shown in Figure~\ref{fig:agn-alone} (and Figset~2), which indicate 
that a galaxy component is almost always indispensable (also see Table~\ref{tbl-3} and the appendix 
Section~\ref{sect:agn-alone}). 

We visually inspect our fits and find that most of the fits are reasonable, with one example shown in 
the left panel of Figure~\ref{fig:decomposition} (for this fit, the reduced chi-squared 
$\chi^2_\nu=3.028$, also indicating that the goodness-of-fit is acceptable). However, for four 
sources, the fits may not be reliable. 
One such example is shown in the right panel of Figure~\ref{fig:decomposition}. For \texttt{PG1626+554},
the FIR data are of low quality and not well-sampled, with only two data points at wavelength 
$>100\ \mu m$. Therefore, it is quite difficult to determine the FIR profile for this source. 
Meanwhile, the reduced chi-squared $\chi^2_\nu$ is $31.973$ which also indicates that the fit is 
poor. We reject all these four sources. Therefore, our final sample for subsequent analysis consists of 
$42$ PG quasars. For the $42$ sources, the $25$-th, $50$-th and $75$-th percentiles of the 
distribution of $\chi^2_\nu$ are $2.55$, $3.30$ and $4.65$, respectively. Six sources have $\chi^2_\nu 
> 10$ because the best-fitting models cannot fit the emission around $6\ \mathrm{\mu m}$ well. Therefore, 
we can keep these six sources since we focus on the AGN SED at much longer wavelengths.

\section{Results and Discussion}
\label{sect:result}
\subsection{Intrinsic AGN IR SED}
\label{sect:sed}
\begin{figure}
\epsscale{1.1}
\plotone{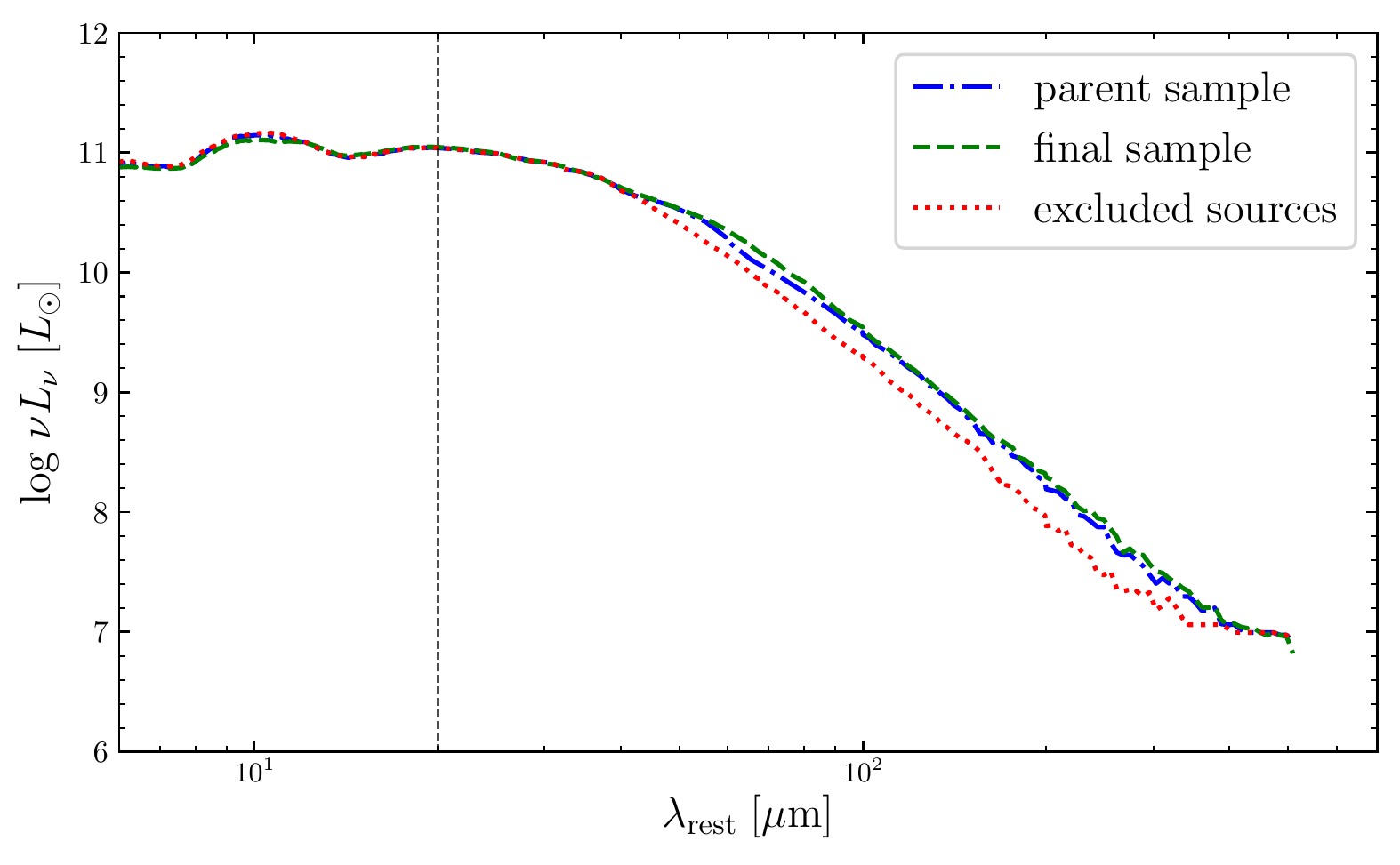}
\caption{The comparison between the median SED of our final sample and that of the rejected sources.
The median SED of the parent sample is also presented. The three median SEDs are consistent with
each other within uncertainties (not shown for clarity).}
\label{fig:bad_good}
\end{figure}

We can now derive the median AGN IR SED from our best-fitting results. \cite{Lani2017} highlighted 
that it is important to normalize the individual SEDs to properly account for a small number of very
FIR-luminous quasars before calculating the median SED. In contrast, S16 directly calculated the 
average AGN IR SED without normalizing the individual SEDs. We construct our median SEDs following 
these two different procedures and find that the two median SEDs are quite similar. We also compare the 
median SED of all the $27$ rejected sources with that of our final sample (Figure~\ref{fig:bad_good}), 
finding that the median SEDs are consistent with each other within uncertainties. In our subsequent 
analysis, we obtain the median SED by first normalizing the individual SEDs to the $40\ \rm{\mum}$
luminosity\footnote{Our conclusions would remain unchanged if we choose to normalize all SEDs to the 
$20\ \rm{\mum}$ luminosity of the S16 AGN mean SED.} of the S16 AGN mean SED and only considering 
the final sample. To assess the differences between two SEDs, we introduce the chi-squared 
$\chi^2=\sum\frac{(L_1-L_2)^2}{dL_1^2+dL_2^2}$ and far-IR luminosity difference $\Delta \log L= 
\mathrm{median} \big\{\left| \log L_1(\lambda>100\ \mathrm{\mu m}) - \log L_2(\lambda>100\ \mathrm{\mu m}) 
\right| \big\}$ as the indicators, where $L_1$ ($L_2$) and $dL_1$ ($dL_2$) represent the $\nu L_\nu$ 
values of the first (second) SED and the corresponding $1\sigma$ uncertainties, respectively. To quantify the 
SED shape, we define 
an IR Color Index (hereafter IRCI) as follows. First, we shift the SEDs into the observed frame by using 
$z=0.088$ (i.e., the median redshift of our sample); second, we calculate the median values of $L_\nu$ 
in the wavelength ranges of \big[$19.89$ \mum, $30.94$ \mum 
\big] (corresponding to the the MIPS $24$ \mum band on \textit{Spizter}; hereafter $L24$) and 
\big[$196.54$ \mum, $298.13$ \mum \big] (i.e., the SPIRE 
$250$ \mum band on \textit{Herschel}; hereafter $L250$); third, IRCI is defined as $\log (L24/L250)$. 
A larger IRCI indicates a hotter SED and vice versa. 

\begin{figure*}
\epsscale{1.1}
\plotone{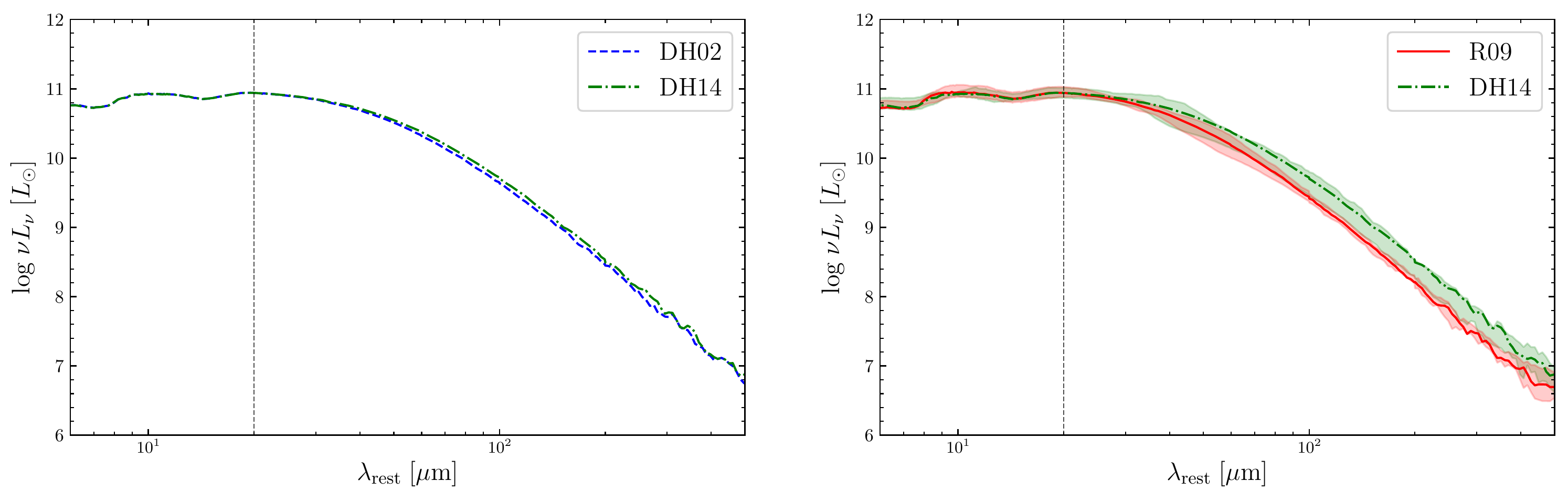}
\caption{The dependence of our median AGN SEDs upon the choice of galaxy templates. Left panel:
the median SED derived with the DH02 galaxy template library versus that with DH14. Right panel:
the median SED derived with DH14 versus that with R09. The shaded regions indicate the interquartile
ranges (i.e., between the $25^\mathrm{th}$ and $75^\mathrm{th}$ percentiles) that are estimated via 
bootstrapping. The three median SEDs are consistent with each other (within $2\sigma$ uncertainties), 
with $\chi^2_\nu=0.86$ and $\delta=0.22$ dex between the DH14 and R09 results. The difference in 
IRCI is $0.39$ dex. }
\label{fig:three_median}
\end{figure*}

We then explore the dependence of the derived median SED upon the galaxy template library. To do 
so, we calculate the median SED using each of the three galaxy template libraries 
\citep{Dale2002, Dale2014, Rieke2009}; the results are presented in 
Figure~\ref{fig:three_median}. It is evident that the median SED of DH02 and that of DH14 
are almost the same. The median SED of R09 and that of DH14 are consistent within 2 
$\sigma$ uncertainties, with a reduced chi-squared $\chi^2_\nu=0.86$ and $\Delta \log L=0.22$ dex; the difference 
between the IRCI (i.e., $\Delta \mathrm{IRCI}$) of the median SED of R09 (IRCI=$3.95$ dex) and that 
of DH14 (IRCI=$3.56$ dex) is $0.39$ dex. These differences can be regarded as the intrinsic 
difference due to different galaxy templates used. When comparing with previous works, we may choose 
any of the three SEDs as long as the intrinsic difference, $\Delta \mathrm{IRCI}=0.39$ 
dex, is taken into account. The median 
SED using the R09 template library and its $1\sigma$ uncertainty is presented in Table~\ref{tbl-2}. 
In subsequent sections, we focus on the median AGN SED derived using the R09 galaxy template library. 

\begin{table*}
\caption{Our AGN median SEDs}
\label{tbl-2}
\begin{tabular}{cccccccccc}
\tableline
Wavelength & All-Median & All-upper & All-lower & hi-Median & hi-upper & hi-lower & lo-Median & lo-upper 
& lo-lower \\
$\mu$m&$\rm log(L_\odot)$&$\rm log(L_\odot)$&$\rm log(L_\odot)$&$\rm log(L_\odot)$&$\rm log(L_\odot)$&$\rm 
log(L_\odot)$&$\rm log(L_\odot)$&$\rm log(L_\odot)$&$\rm log(L_\odot)$\\
(1)        & (2)    & (3)          & (4)          & (5)    & (6)      & (7)      & (8)    & (9)      & (10)     
\\
\tableline
\tableline
6.0 & 10.72 & 10.838 & 10.691 & 10.812 & 10.847 & 10.634 & 10.67 & 10.806 & 10.558 \\
6.1 & 10.726 & 10.846 & 10.695 & 10.808 & 10.837 & 10.62 & 10.671 & 10.803 & 10.566 \\
6.2 & 10.726 & 10.846 & 10.703 & 10.811 & 10.848 & 10.638 & 10.675 & 10.806 & 10.56 \\
6.3 & 10.736 & 10.858 & 10.703 & 10.813 & 10.848 & 10.638 & 10.674 & 10.807 & 10.555 \\
6.4 & 10.729 & 10.853 & 10.705 & 10.804 & 10.838 & 10.627 & 10.681 & 10.798 & 10.585 \\
6.6 & 10.725 & 10.859 & 10.704 & 10.806 & 10.85 & 10.627 & 10.685 & 10.798 & 10.593 \\
6.8 & 10.716 & 10.856 & 10.703 & 10.8 & 10.845 & 10.617 & 10.692 & 10.788 & 10.6 \\
6.9 & 10.72 & 10.857 & 10.703 & 10.797 & 10.843 & 10.626 & 10.687 & 10.79 & 10.61 \\
7.0 & 10.717 & 10.856 & 10.703 & 10.788 & 10.823 & 10.622 & 10.689 & 10.783 & 10.598 \\
7.1 & 10.719 & 10.858 & 10.709 & 10.797 & 10.839 & 10.617 & 10.7 & 10.792 & 10.637 \\
7.3 & 10.72 & 10.861 & 10.712 & 10.789 & 10.834 & 10.625 & 10.704 & 10.784 & 10.637 \\
7.5 & 10.73 & 10.865 & 10.729 & 10.791 & 10.831 & 10.617 & 10.717 & 10.786 & 10.662 \\
7.6 & 10.741 & 10.866 & 10.736 & 10.8 & 10.848 & 10.636 & 10.718 & 10.795 & 10.652 \\
7.7 & 10.756 & 10.877 & 10.741 & 10.812 & 10.859 & 10.658 & 10.736 & 10.811 & 10.674 \\
7.8 & 10.774 & 10.884 & 10.743 & 10.819 & 10.858 & 10.663 & 10.733 & 10.815 & 10.657 \\
7.9 & 10.808 & 10.895 & 10.761 & 10.843 & 10.886 & 10.697 & 10.747 & 10.837 & 10.682 \\
8.0 & 10.833 & 10.907 & 10.772 & 10.866 & 10.892 & 10.719 & 10.767 & 10.856 & 10.707 \\
8.2 & 10.868 & 10.932 & 10.79 & 10.905 & 10.916 & 10.769 & 10.787 & 10.88 & 10.728 \\
8.4 & 10.901 & 10.965 & 10.802 & 10.923 & 10.936 & 10.809 & 10.81 & 10.902 & 10.758 \\
8.5 & 10.908 & 10.984 & 10.805 & 10.935 & 10.953 & 10.826 & 10.816 & 10.907 & 10.767 \\
\tableline
\end{tabular}
\tablecomments{(1) Wavelength; (2)--(10) $\mathrm{log}\ \nu L_\nu$ of our AGN SEDs; 
``hi-Median'' and ``lo-Median'' refer to the median SEDs of the $\mathrm{log}\ L_{5100}>44.55$ and 
the $\mathrm{log}\ L_{5100}<44.55$ subsamples, respectively; ``-upper'' and ``-lower'' postfixes refer 
to the $75^{\mathrm{th}}$- and $25^{\mathrm{th}}$-percentile SEDs for each sample, which are 
derived by bootstrapping. The SEDs are normalized to the S16 average AGN SED at $40\ \mathrm{\mu m}$. 
The full table is published in its entirety in the machine-readable format. A portion is 
shown here for guidance purpose. 
}
\end{table*}

\begin{figure*}
\epsscale{1.1}
\plotone{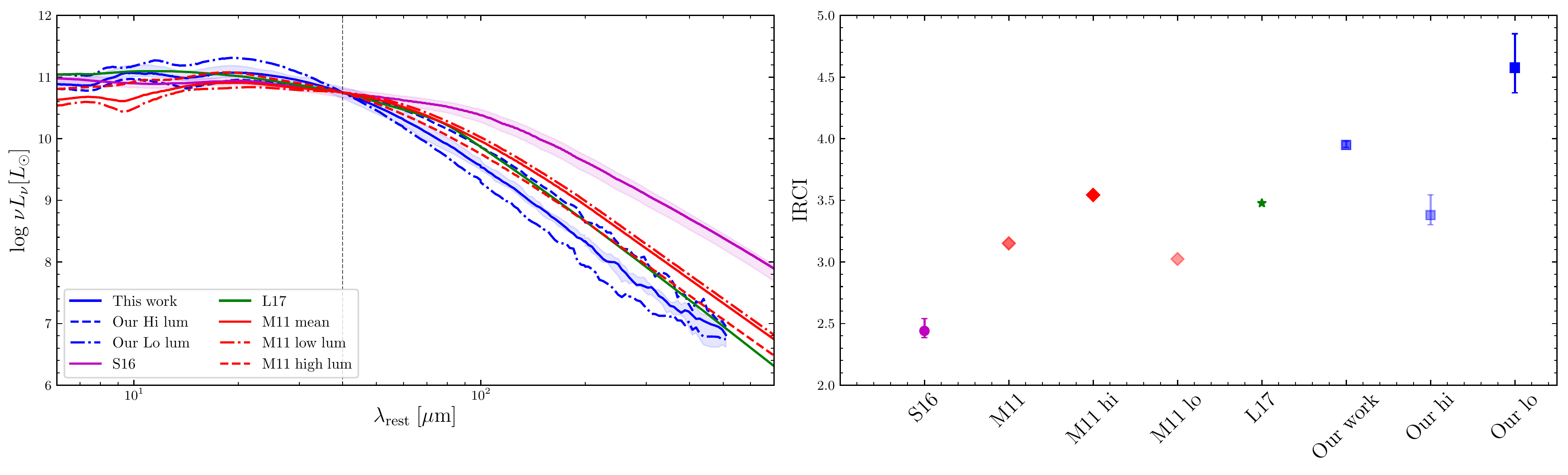}
\caption{A comparison between our results and previous works. The blue solid curve represents 
the median AGN IR SEDs of our full sample. And the pink, green, and red 
curves are from S16, L17, and M11. The dashed and dash-dotted curves indicate the median AGN SEDs 
of high- and low-luminosity sample in this work, respectively. Our median AGN IR SED is roughly 
consistent with that of L17 and the M11 high-luminosity AGNs within uncertainties, but significantly 
disagree with that of S16. The shaded regions represent available 1~$\sigma$ uncertainties (derived 
via bootstrapping).}
\label{fig:final}
\end{figure*}

\subsection{The IR SED as a function of AGN luminosity}
\label{sect:SED_lum}
We now compare our median AGN SED with previous works (Figure~\ref{fig:final}). The IRCIs 
for our result and previous works are shown in the right panel of Figure~\ref{fig:final}. Our AGN 
median IR SED is only slightly hotter than that of L17 with $\Delta \mathrm{IRCI}=0.47$ dex (i.e., 
close to the intrinsic scatter due to using different galaxy templates). Our median AGN SED is also 
somewhat hotter than that of M11 (with $\Delta \mathrm{IRCI}=0.80$ dex, and $\Delta \log L= 0.57$ dex). The 
AGNs adopted by M11 are less luminous than those of our final sample, since they used 
moderate-luminosity AGNs with a median bolometric luminosity $L_{\rm{bol}}=10^{44.05}\;\rm{erg\ 
s^{-1}}$ calculated using the \citet{Lusso2012} bolometric-correction relation, in contrast to our 
PG quasars with a median $L_{\rm{bol}}=10^{45.55}\ \rm{erg\ s^{-1}}$. It is worth noting that our 
median AGN IR SED is close to the M11 high-luminosity AGN median SED (i.e., $\Delta 
\mathrm{IRCI}= 0.40$ dex and $\delta=0.30$ dex.) Therefore, we speculate that the differences between 
our median AGN SED and that of M11 are caused by two factors: first, the AGN IR SED of 
more luminous AGNs is hotter than that of the less luminous ones; and second, the data of M11 have 
limited FIR coverage and their SED is biased (see Section~\ref{sect:SED_lum}). 

\begin{figure*}
\epsscale{1.1}
\plotone{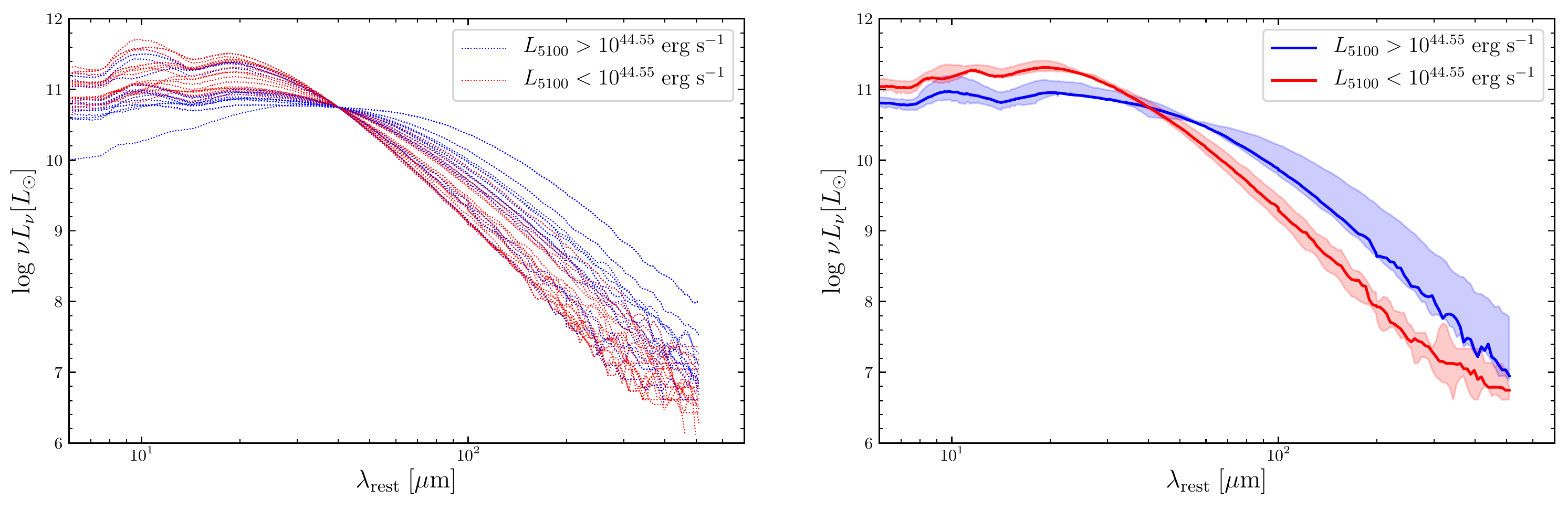}
\caption{The AGN SEDs for our high- and low-luminosity subsamples. Left panel: Blue (red) dotted curves 
represent AGN SEDs of sources with $L_{5100}$ greater (less) than the median $L_{5100}$ ($10^{44.55}\ 
\mathrm{erg\ s^{-1}}$). Right panel:The blue and red solid curves represent the median SEDs of the high- 
and low-luminosity AGN subsamples, respectively. Shaded regions are the 1 $\sigma$ uncertainties derived 
by bootstrapping. Luminous sources tend to have cooler SEDs (i.e., larger FIR fluxes for fixed MIR 
emission) than less luminous ones.}
\label{fig:Lum_dark}
\end{figure*}

To test our speculations of the differences between our median AGN SED and that of M11, we
explore the IR SED as a function of AGN luminosity. We split the final sample into 
two subsamples, each containing $21$ sources: the 
high-luminosity (i.e., $L_{5100}\geqslant \tilde{L}_{5100}$, where $\tilde{L}_{5100}=10^{44.55} 
\;\rm{erg\;s^{-1}}$ is the median $L_{5100}$ of our full sample) subsample with a median $L_{5100}= 
10^{44.91}\;\rm{erg\;s^{-1}}$ and the low-luminosity subsample ($L_{5100}<\tilde{L}_{5100}$) with 
a median $L_{5100}=10^{44.11}\;\rm{erg\;s^{-1}}$. We construct the median SEDs for the two subsamples 
(Figure~\ref{fig:Lum_dark}) and find that the high-luminosity subsample (IRCI=$3.38$ 
dex) tends to have a cooler median SED (i.e., having systematically higher FIR emission and a smaller 
IRCI) than the low-luminosity subsample (IRCI=$4.58$ dex; therefore, $\Delta \mathrm{IRCI}=1.20$ dex). 
The median $L_{\mathrm{bol}}$ of L17 is $10^{45.85}\ \mathrm{erg\ s^{-1}}$, which is close to 
that of our high-luminosity one. The AGN IR SED of L17 is also more consistent with that of our 
high-luminosity subsample ($\Delta \mathrm{IRCI}=0.10$ dex) than that of our full sample or low-luminosity 
subsample. The AGN IR SED of M11 which is obtained from much less luminous AGNs is cooler (i.e., 
smaller IRCI) than our results. In addition, M11 found that low-luminosity AGNs tend to have cooler SEDs 
than high-luminosity AGNs (with $\Delta \mathrm{IRCI}=-0.52$ dex), which is inconsistent with 
the tendency of our results. This inconsistency might be caused by the limited FIR coverage in M11. 
M11 used the IRAS photometric data that only reach the observed-frame wavelength of 100 $\mum$. They 
adopted a modified blackbody shape of AGN SED at FIR bands, but there were only two FIR data points 
(i.e., 60 and 100 $\mum$ fluxes, most of which are upper limits) to constrain their SED decomposition. 
Whereas, we have at least 4 {\it Spitzer}/{\it Herschel} FIR data points (i.e., 60, 100, 160, and 250 
$\mum$ fluxes) to define our AGN IR SEDs. 

\begin{figure*}
\epsscale{1.1}
\center{}
\includegraphics[width=0.33\linewidth]{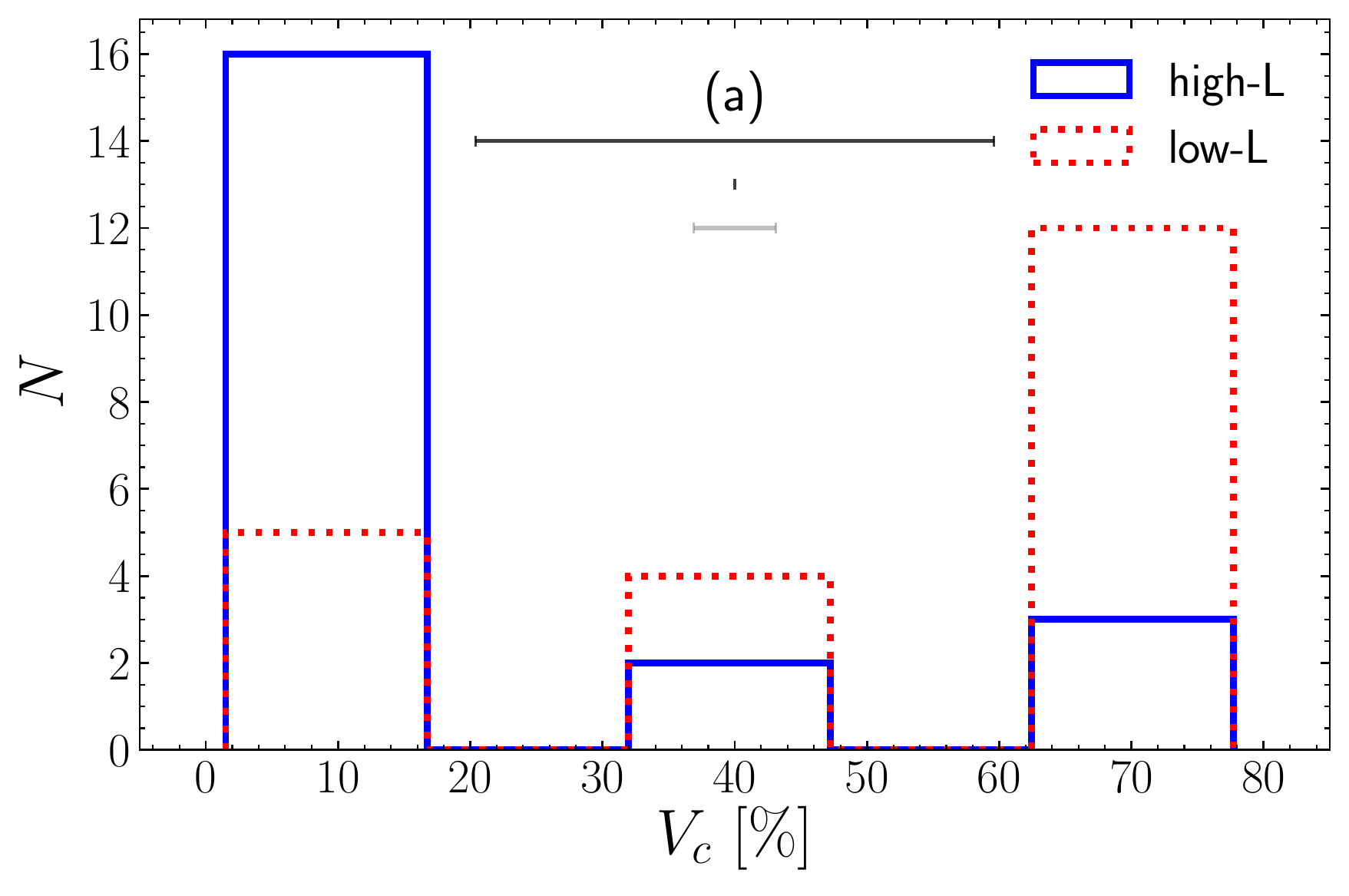}
\includegraphics[width=0.33\linewidth]{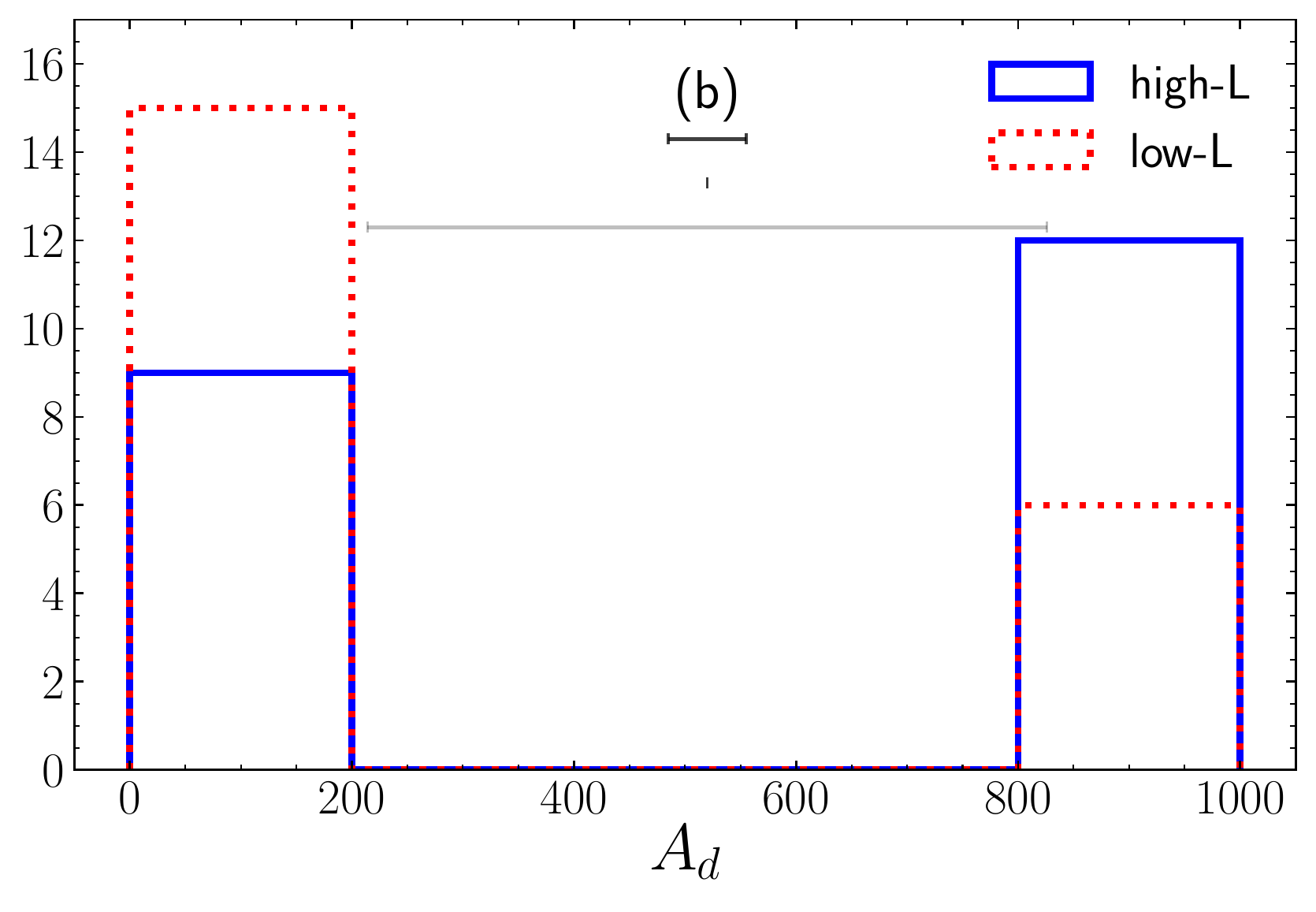}
\includegraphics[width=0.33\linewidth]{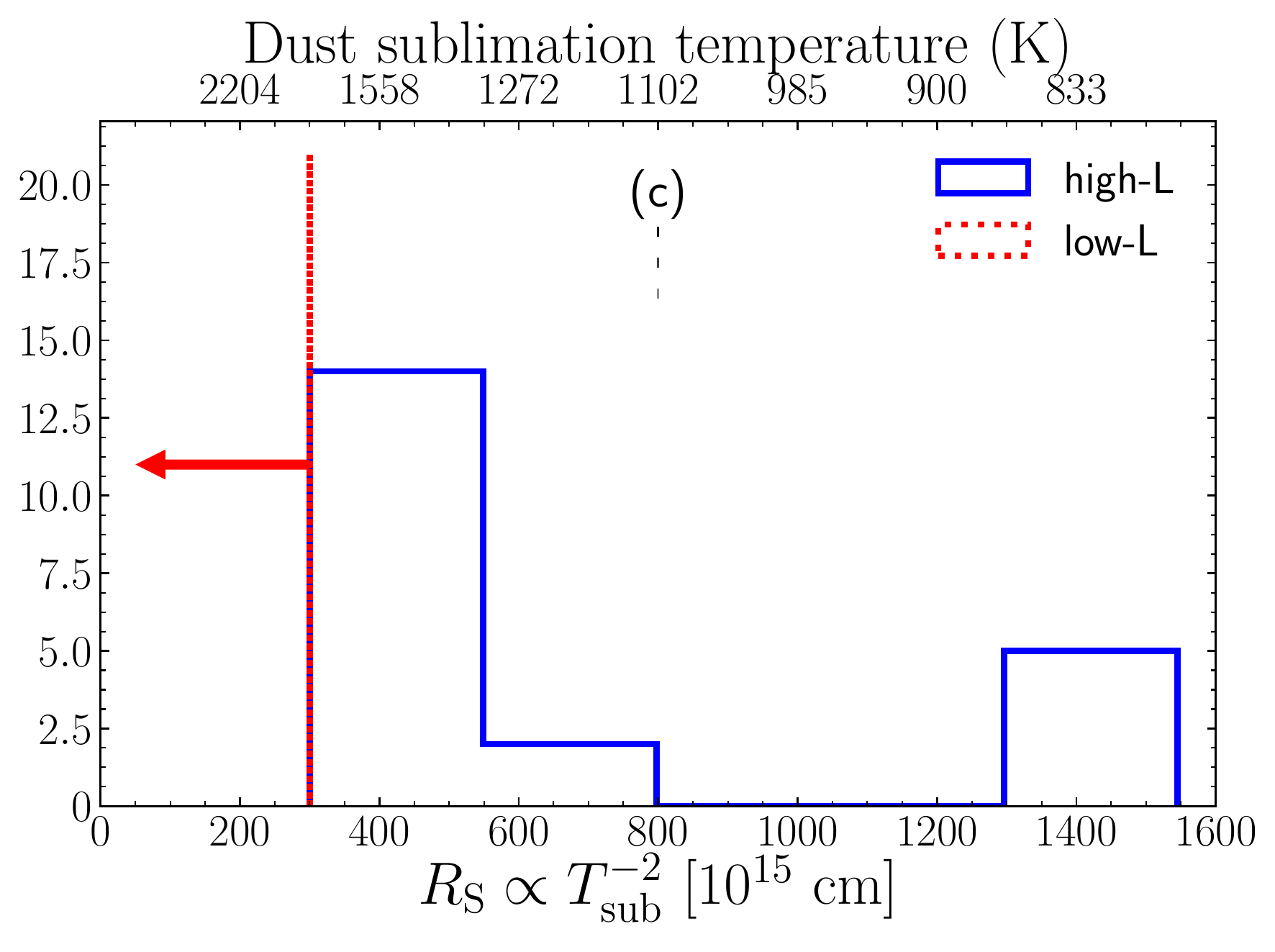}\\
\includegraphics[width=0.33\linewidth]{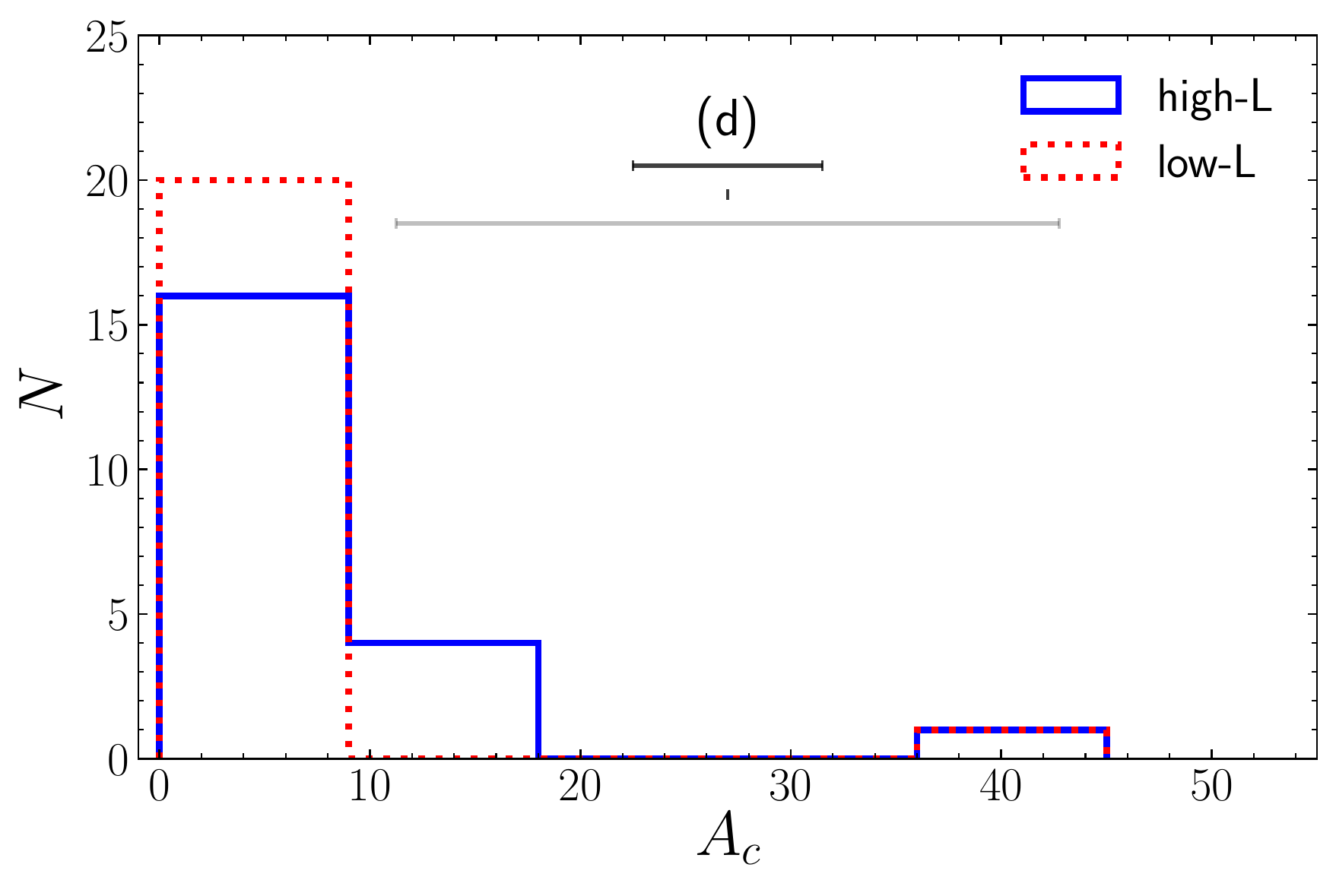}
\includegraphics[width=0.33\linewidth]{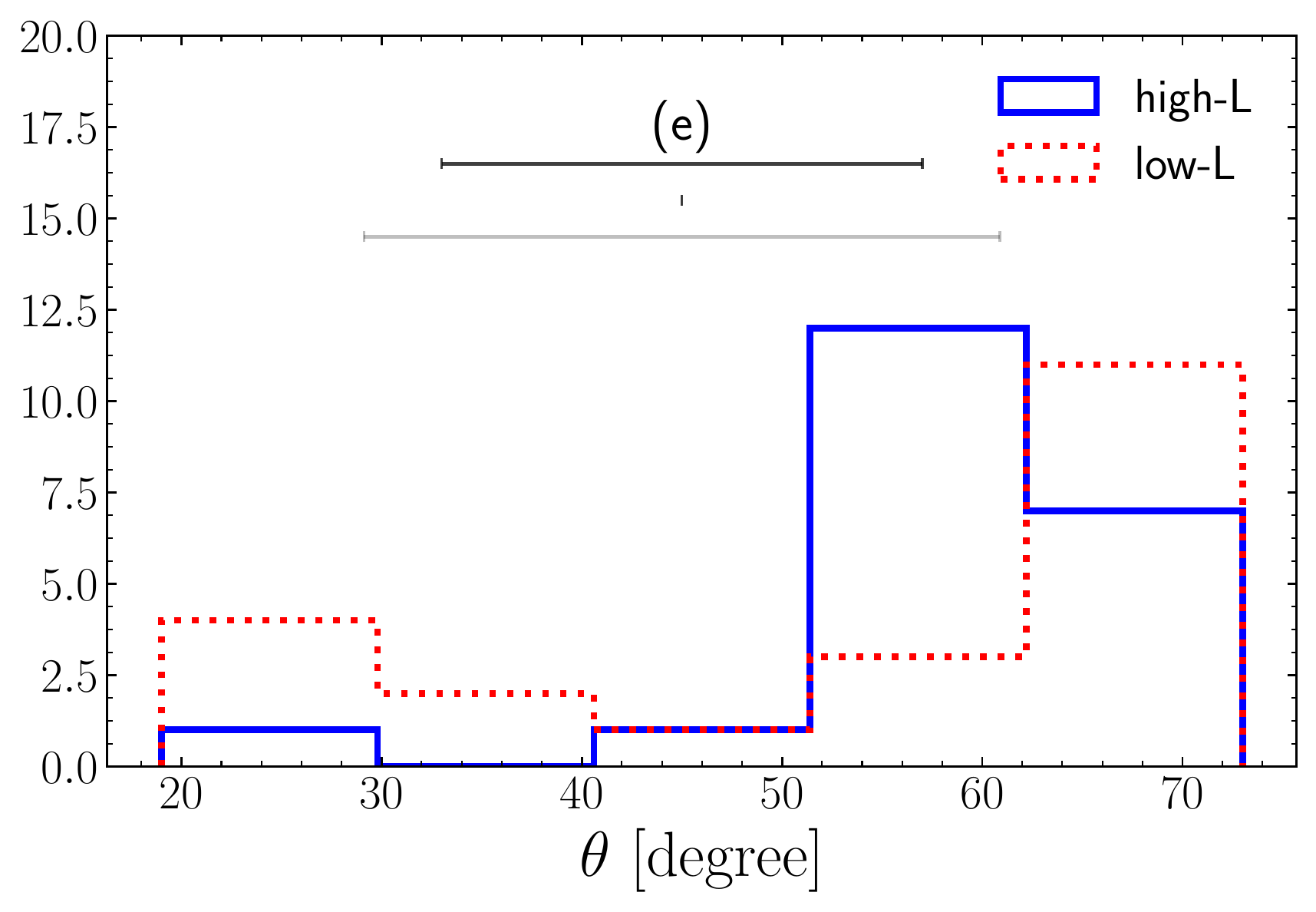}
\caption{The comparison of the distributions of $V_c$, $A_d$, $R_s$, $A_c$ and $\theta$ for the 
high-luminosity subsample (blue histograms) and low-luminosity subsample (red histograms). The 
distribution of $A_d$ shows a gap between $200$ and 800, which might be caused by the large gap in 
the parameter space of $A_d$ (see Table~\ref{tbl-1}) in the model of \cite{Siebenmorgen2015}. 
High-luminosity AGNs tend to have larger mid-plane optical depths ($A_d$), smaller cloud filling factors 
($V_c$), and larger inner radii ($R_s$; which indicates colder dust sublimation temperature). Interestingly, 
the low-luminosity AGNs seem to prefer $R_s$ values that hit the lower bound of the allowed $R_s$ space. 
The distributions of $A_c$ 
(or $\theta$) for the two subsamples are statistically indistinguishable 
given $p>0.05$ from the Anderson-Darling tests. In each panel, the three horizontal lines from bottom 
to top show the corresponding error bars of the three representative sources with the 
$25^\mathrm{th}$-/$50^\mathrm{th}$-/$75^\mathrm{th}$-percentile luminosities in our final sample, respectively. 
Some parameters have zero error bars; this is because the parameter space of the torus model is too 
sparse to infer the $1\sigma$ errors, i.e., the $1\sigma$ errors are smaller than the step sizes of the parameters 
(see Table~\ref{tbl-1}). The leftward red arrow in panel (c) indicates the trend of $R_\mathrm{S}$ for the 
low-luminosity subsample if the $R_\mathrm{S}$ parameter space can reach smaller values.}
\label{fig:three_parameter}
\end{figure*}

The dependence of AGN IR SED on luminosity might be caused by the change of torus properties. To look 
into this, we collect the best-fitting torus parameters from our SED decomposition. In Figure~\ref{fig:three_parameter}, 
we compare the distributions of each parameter between the low-/high-luminosity samples.\footnote{To demonstrate 
the uncertainties of the best-fitting parameters, we choose three representative sources whose luminosities correspond 
to the $25^\mathrm{th}$, $50^\mathrm{th}$, and 
$75^\mathrm{th}$ percentile luminosity in our sample to represent the low-/median-/high-luminosity sources. 
First, we create 128 mock SEDs for each of the three sources by adding Gaussian noise to the observed SEDs. 
For each source, we then fit mock SEDs following the same methodology and obtain the distributions of the 
torus parameters. We report the differences between the $16^{\rm th}$ and $84^{\rm th}$ percentiles of each distribution 
as the error bar for each parameter (i.e., the horizontal lines in each panel of Figure~\ref{fig:three_parameter}).} 
We find that high-luminosity AGNs tend to have larger mid-plane optical depths and smaller cloud filling 
factors than low-luminosity ones (see Figure~\ref{fig:three_parameter}; indeed, 
in each panel, the Anderson-Darling test indicates that, at the 99\% significance level or above, the 
null hypothesis that the two distributions are the same can be rejected). It is also clear in the right 
panel that luminous AGNs have tori of larger inner radii $R_s$. 

In the torus model of \citet[][see their Section 2.2]{Siebenmorgen2015}, the AGN 
luminosity is fixed to be $10^{11} L_{\odot}$ and $R_S$ is allowed to vary. Then, to understand the meaning of 
$R_S$, let us consider a dusty torus model with fixed dust composition whose geometric distribution is 
sublimation-radius-scale-invariant. The only 
variable parameters are the dust sublimation radius $r_{\mathrm{sub}}$ (which is not $R_S$) and the AGN luminosity 
($L_{\mathrm{AGN}}$). The AGN flux $F_{\mathrm{AGN}}$ received at the sublimation radius, i.e., $F_{\mathrm{AGN}} = 
L_{\mathrm{AGN}}/(4\pi r^2_{\mathrm{sub}})$, should equal the flux of the re-emitted infrared blackbody 
emission (i.e., $\sigma T_{\mathrm{sub}}^4$, where $\sigma$ and $T_{\mathrm{sub}}$ are the Stefan–Boltzmann 
constant and the dust sublimation temperature, respectively) under the assumption of steady state. Therefore, 
it is evident that 
\begin{equation}
\label{eq:Tsub}
r_{\mathrm{sub}} = R_S\sqrt{L_{\mathrm{AGN},0}/L_{\mathrm{AGN}}} \\,
\end{equation}
where $R_S=(L_{\mathrm{AGN},0}/(4\pi\sigma T_{\mathrm{sub}}^4))^{0.5}$ is the sublimation radius 
for an AGN with $L_{\mathrm{AGN},0}=10^{11} L_{\odot}$ whose torus dust sublimation temperature is identical 
to the AGN with $L_{\mathrm{AGN}}$. That is, the $L_{\mathrm{AGN}}\propto r^2_{\mathrm{d}}$ relation \citep[which 
has been observed; see, e.g.,][]{Kishimoto2011} adopted by \cite{Siebenmorgen2015} ensures that an AGN with 
$L_{\mathrm{AGN}}$ and $r_{\mathrm{sub}}$ shares the same dust sublimation temperature with another AGN with 
$L_{\mathrm{AGN},0}$ and $R_S$. Then, the two AGNs have the same torus SED shape.\footnote{This 
is why the AGN luminosity is fixed to be $10^{11} L_{\odot}$ in \citet[][see their Section 2.2]{Siebenmorgen2015}.} 
We expect that, since \cite{Siebenmorgen2015} fixed the AGN luminosity to be $L_{\mathrm{AGN},0}=10^{11} L_{\odot}$, 
$R_S \propto T_{\mathrm{sub}}^{-2}$, i.e., $R_S$ actually indicates the dust sublimation temperature (as indicated 
in Figure~\ref{fig:three_parameter} (c)). Therefore, our result that luminous AGNs tend to have larger $R_S$ suggests 
that higher-luminosity AGNs tend to have lower dust sublimation temperatures. 

It should be noted that the torus model of \cite{Siebenmorgen2015} is a phenomenological model. 
The physical reasons for the differences between the IR SEDs of the high- and low-luminosity subsamples can be 
very complicated. We suspect that, compared with the low-luminosity subsample, AGNs in the high-luminosity 
subsample might have smaller covering factors (see panel (a) of Figure~\ref{fig:three_parameter}; possibly because 
more luminous AGNs can more effectively swipe gas and dust away than the less luminous ones) and the UV/optical 
photons from the central engine can heat the 
dust on the galaxy scale \citep{Sanders1989}; these dust clouds are presumably colder than the dusty torus. As a result, 
the IR SED of 
the high-luminosity subsample is cooler than the low-luminosity one. Moreover, the high-luminosity sources presumably 
harbor more massive supermassive black holes than the low-luminosity counterparts. Then, it is natural to expect 
that the host galaxies of the high-luminosity sources are also more massive and have higher metallicity than those 
of the low-luminosity ones. The difference in metallicity might also be responsible for the differences in IR 
SEDs \cite[e.g.,][]{Engelbracht2008}. A detail investigation of the physical mechanisms that are responsible for 
the IR SED differences is beyond the scope of this work. 

M11 argued that higher-luminosity AGNs can heat a larger fraction of their surrounding dust to higher 
temperatures and they tend to produce a warmer SED (i.e., having stronger emission at MIR wavelengths). 
In this work, we show that higher-luminosity AGNs tend to have lower dust temperatures. 
Therefore, higher-luminosity AGNs tend to show more FIR emission 
with respect to MIR emission and have smaller IR color indices \citep[see also Section~3 and Figure~5 
of][]{Siebenmorgen2015}. Therefore, as the AGN luminosity increases, the SED becomes cooler.

It should be noted that the AGN IR SED of S16 is luminosity-independent (see Figure~6 of S16). The lack of
luminosity-dependency might indicate that the stellar contribution is not well subtracted from the total SED
(see Section~\ref{sect:discussion} for more discussions).

\subsection{Reliability of PAH relating to galaxy infrared properties}
\label{sect:discussion}
\begin{figure}
\epsscale{1.1}
\plotone{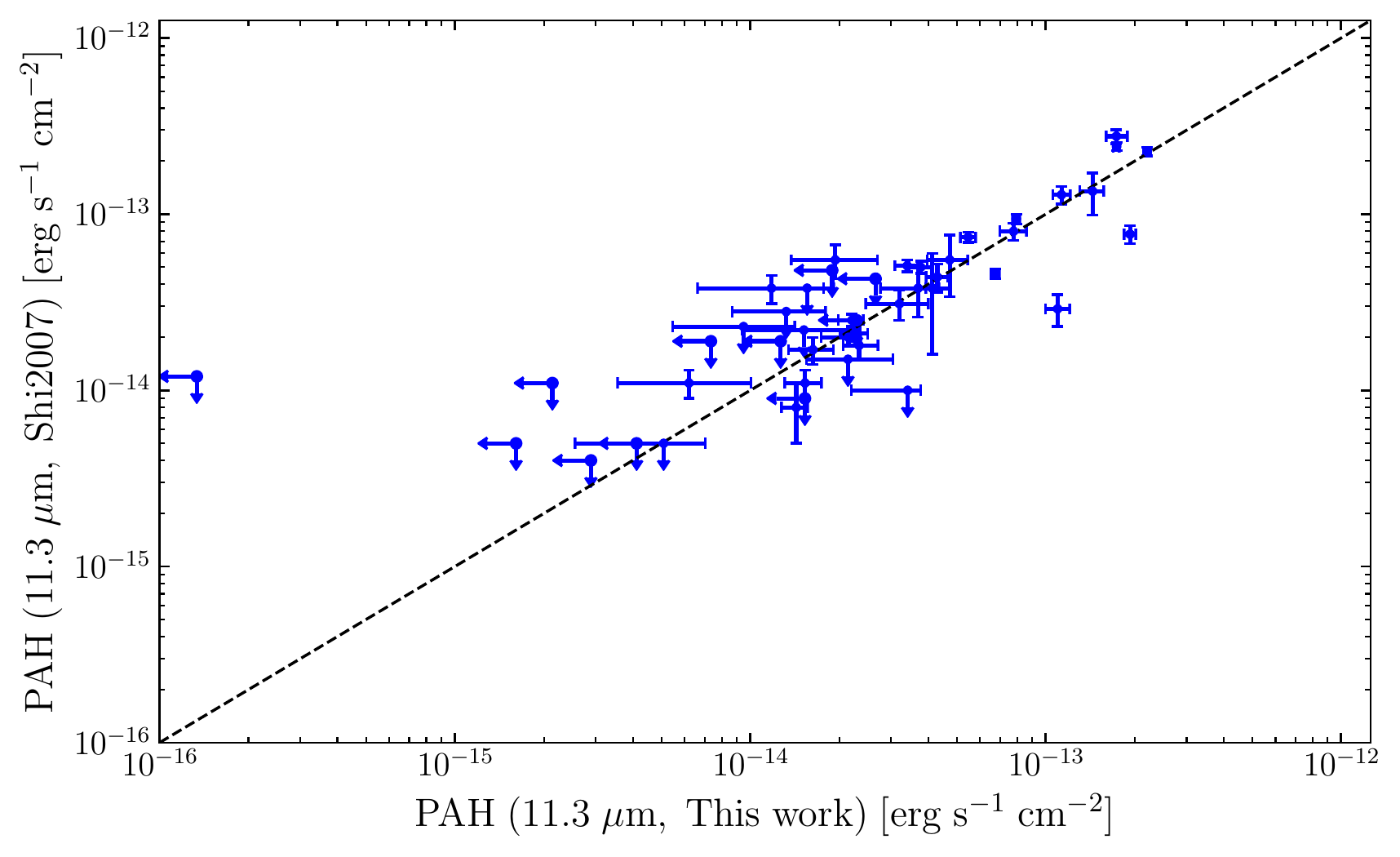}
\caption{The comparison of $11.3\ \mum$ PAH flux measurements of the observed IRS data between our work
and \cite{Shi2007}. Error bars indicate the $25^\mathrm{th}$- and $75^\mathrm{th}$- percentiles; 
upper limits correspond to the $90$-th percentiles. The dashed line indicates the one-to-one 
relation. There is no significant systematic offset between the two sets of measurements.}
\label{fig:pah_07_14}
\end{figure}

S16 used PAH fluxes and the DH02 galaxy template library to determine the stellar contribution to the total 
SED. However, the PAH features can be easily contaminated by the nearby Silicate absorption. 
To verify the reliability of using the $11.3\ \mu$m PAH strength as an indicator of the FIR emission 
due to star formation, we check the PAH fluxes of our best-fitting models 
for the three galaxy libraries 
by adopting the following methodology for the S16 sample.\footnote{All but one S16 
sources are also in our parent sample. Therefore, we can use our SED decomposition method (see 
Section~\ref{sect:method}) to find the best-fitting AGN and galaxy templates for the S16 sources.} First, 
we simulate mock IRS spectra according to our 
best-fitting AGN and galaxy templates. During the simulation, we ensure that the mock spectra and the observed 
IRS spectra have the same wavelength resolution and suffer from the same measurement errors. We then fit 
both the mock spectra and the observed IRS data following the same methodology of \cite{Shi2007}. 
We choose to fit the observed IRS data rather than use the result of \cite{Shi2007} because the observed IRS 
data have been re-calibrated and re-processed since \cite{Shi2007}. To estimate the measurement 
uncertainties on PAH fluxes, we adopt a Monte Carlo approach by adding Gaussian flux-density noise 
(with a standard deviation being the measurement error) to the flux in each bin of wavelength and 
generating a new spectrum. We then refit the new spectrum following the same methodology. For each source, 
this procedure is repeated $1000$ times to obtain the distribution of the PAH flux; the measurement uncertainties 
are then estimated as the $25^\mathrm{th}$- and $75^\mathrm{th}$- percentiles of the PAH flux distribution. 
For some sources, the $25^\mathrm{th}$ percentiles are smaller than zero; we report the $90$th percentiles 
as the upper limits. A comparison between our measurements of the IRS data and those of \cite{Shi2007} is 
illustrated in Figure~\ref{fig:pah_07_14}. Overall, we find no significant systematic offset between the two 
sets of measurements albeit with some scatter. This is not surprising given that the fitting method 
is the same but the data calibration has been updated \citep{Shi2014}. As for sources without 
the detections of the PAH $11.3\ \mu$ features, \cite{Shi2007} used the $5\sigma$ measurement noise 
as the upper limits; instead, we report the $90^{\mathrm{th}}$-percentiles of the Monte Carlo simulations as 
the upper limits. Therefore, at the low flux end (i.e., $<3\times10^{-14}\ 
\mathrm{erg\ s^{-1}cm^{-2}}$), the results of \cite{Shi2007} are larger than our measurements by (on average) 
a factor of $\sim 2$. 

\begin{figure*}
\epsscale{1.1}
\plotone{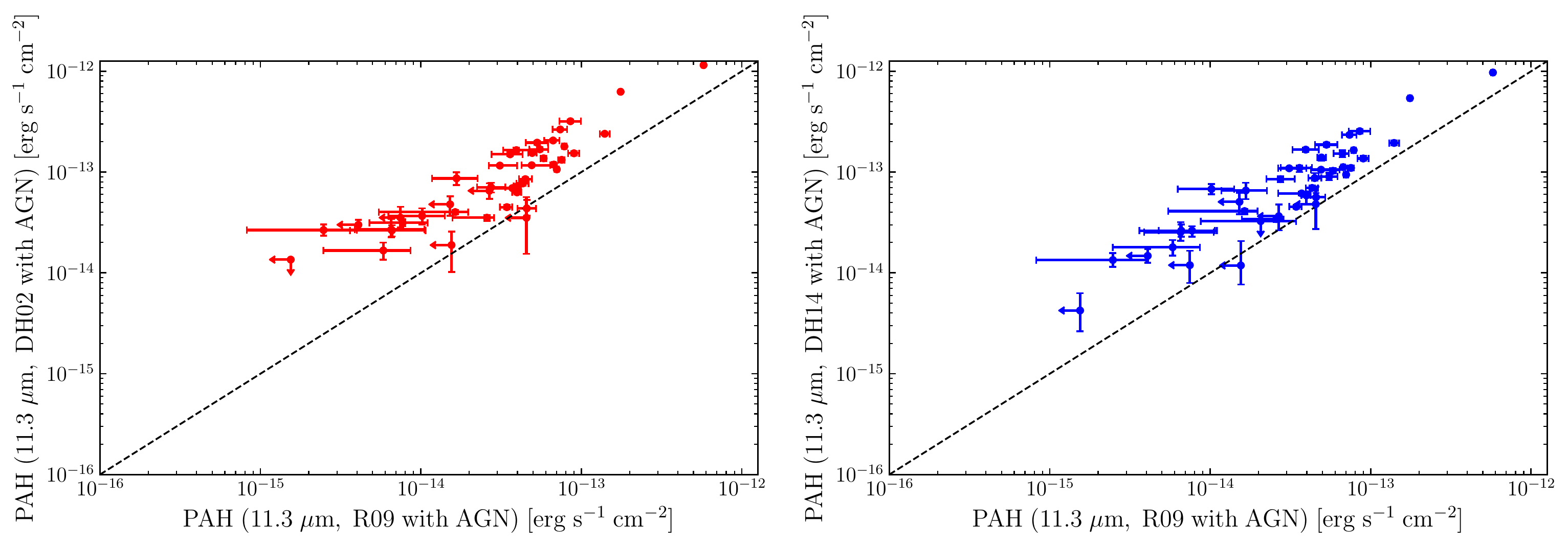}
\caption{The measurements of $11.3\ \mum$ PAH fluxes of the mock spectra using different galaxy template
libraries. Left panel: The result of DH02 mock spectra versus that of R09 mock spectra. Right panel: The
result of DH14 mock spectra versus that of R09 mock spectra. The dashed lines indicate the one-to-one
relations. The DH02 and DH14 results are systematically larger than the R09 result. Compared to the R09
result, for the fixed PAH flux measured from real data, the DH02 and DH14 libraries underestimate the
galaxy FIR fluxes (see section \ref{sect:discussion}).}
\label{fig:pah_three_comparison}
\end{figure*}

The PAH fluxes of mock spectra for the three galaxy template libraries are presented in 
Figure~\ref{fig:pah_three_comparison}. It is clear that the PAH fluxes derived using the DH02 and DH14 
libraries are systematically larger than those derived with R09. Indeed, the median PAH strength 
values of DH02 and DH14 are 181\% and 165\% than those of R09, respectively. This 
difference is likely a result of contamination by the silicate absorption, which is not accounted for  
in DH02 or DH14 but taken into account in R09. If the silicate absorption is indeed important, the observed 
PAH flux would be an under-estimation of the true value. In addition, if one estimates the galaxy contribution 
to the total SED by selecting the DH02 or DH14 template with the PAH flux that is closest to the 
observed one (e.g., S16), the galaxy contribution would also be underestimated; this is not the case if 
one adopts the R09 library \citep[see also Figure 3 of][]{Lyu2017}. Therefore, using the PAH strength to 
determine the galaxy contribution to the total SED is highly uncertain and template-dependent. M11 and 
our work derive the AGN SED by performing SED decomposition and do not suffer this problem. 

\begin{figure}
\epsscale{1.1}
\plotone{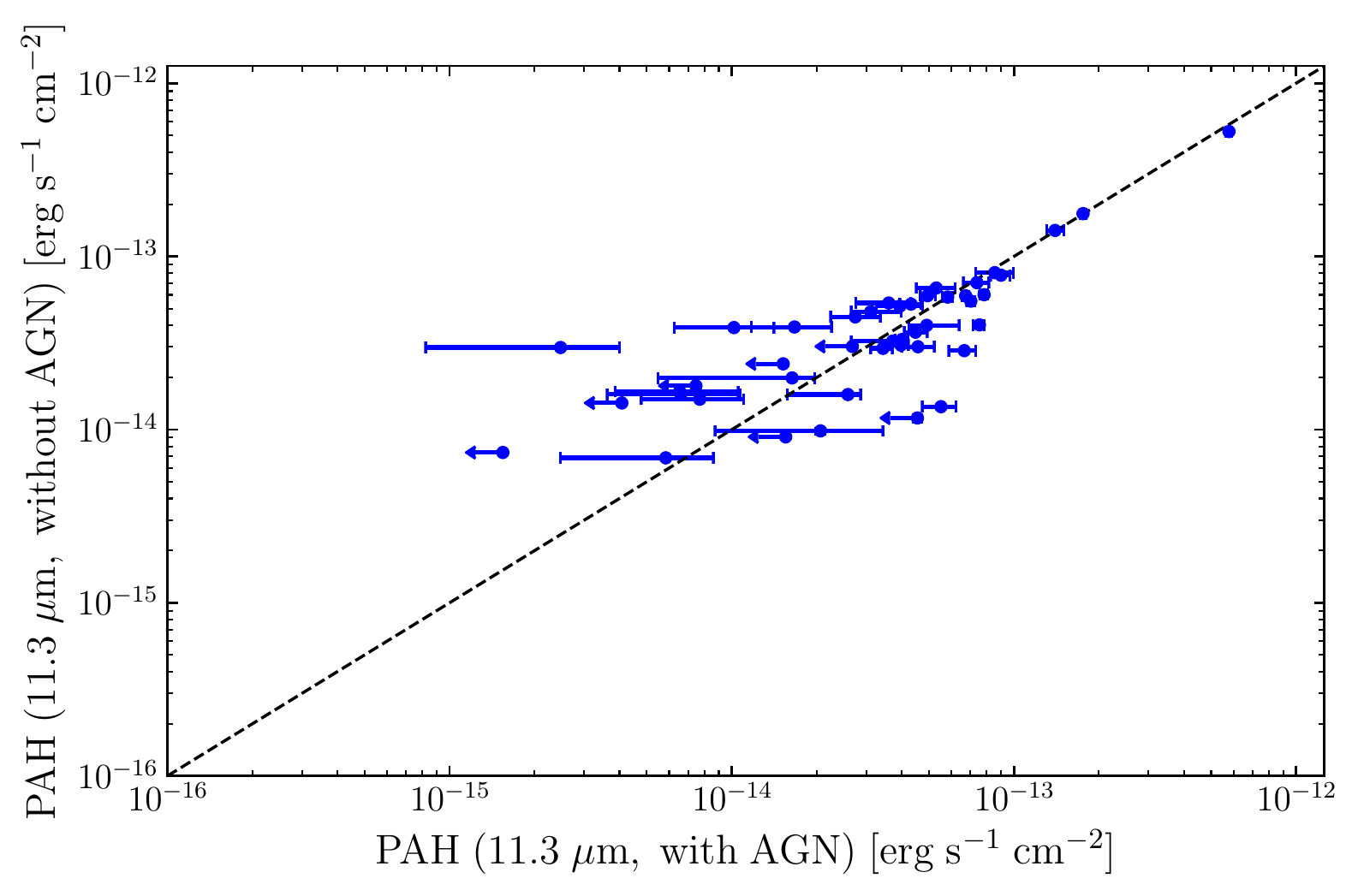}
\caption{The comparison between $11.3\ \mum$ PAH flux measurements for R09 mock spectra with and
without the AGN component. The dashed line indicates the one-to-one relation. In the presence 
of AGN contamination, the PAH flux is slightly underestimated. 
The dispersion can be large when the PAH feature is very weak.}
\label{fig:AGN_effect_PAH}
\end{figure}

The MIR spectra are often dominated by AGN emission. The PAH signal is often very weak and diluted 
by the silicate absorption feature in an AGN spectrum. Therefore, the PAH flux measurement 
might be underestimated.
To illustrate this possible underestimation, we generate two sets of mock IRS spectra: 
for each source in our sample, we generate a mock spectrum by only considering the best-fitting 
R09 template and another mock spectrum by combining the best-fitting R09 and AGN templates (i.e., according 
to Eq.~\ref{eq:f1}). Then, we measure the PAH fluxes from the two sets of mock spectra 
(Figure~\ref{fig:AGN_effect_PAH}). In the presence of AGN contamination, the PAH flux is overall slightly 
underestimated (i.e., by 20\%), and this bias is more evident towards the low PAH flux end.

\begin{figure}
\epsscale{1.1}
\plotone{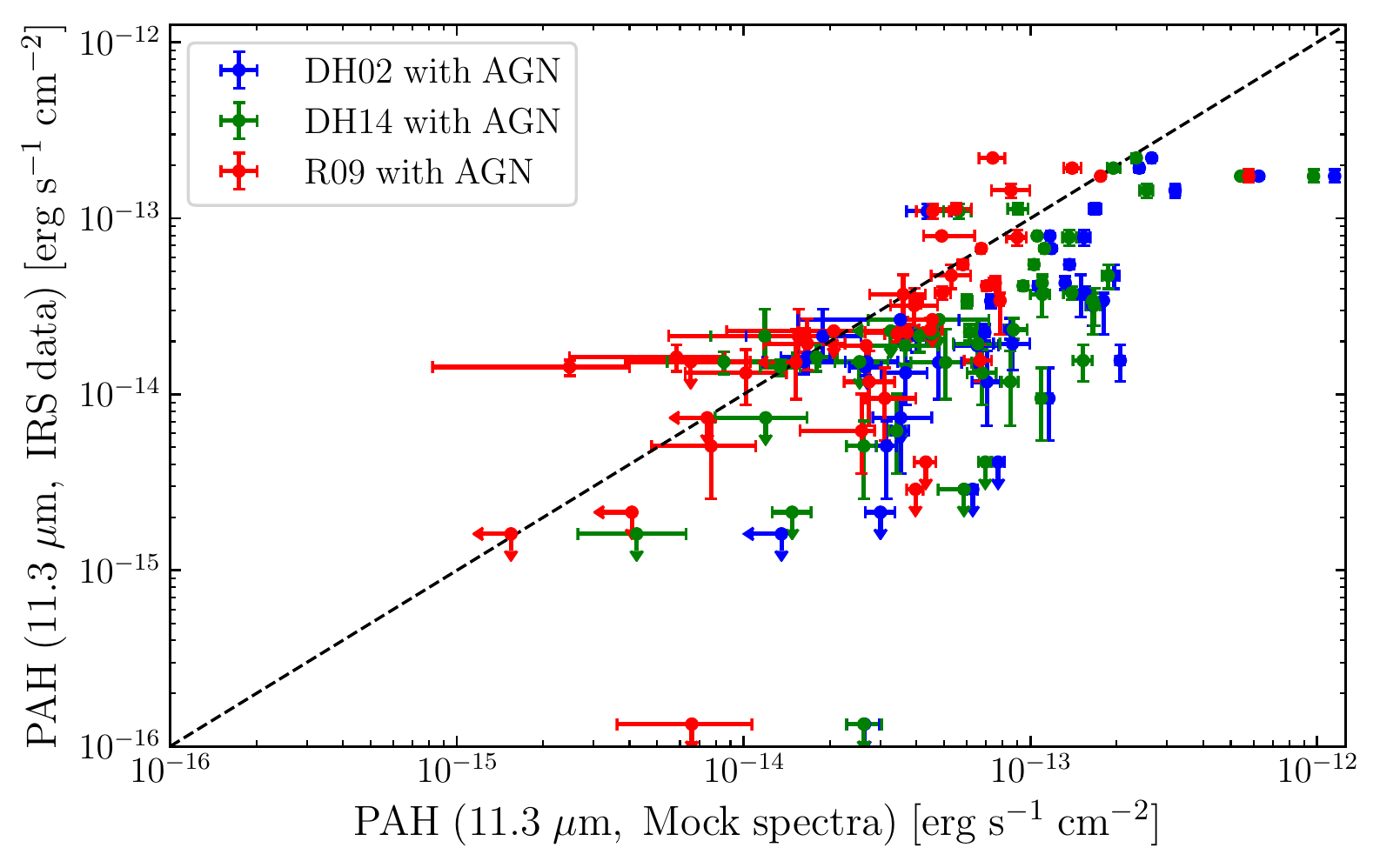}
\caption{The $11.3\ \mum$ PAH fluxes of our mock spectra versus those of the observed IRS data. The red,
blue, and green dots are for the R09, DH02, and DH14 libraries. The dashed line indicates the one-to-one
relation. Among the three galaxy template libraries, the R09 PAH emission is roughly consistent with
observations. For the DH02 and DH14 libraries, the PAH emission is over-estimated. This is likely
due to the fact that the silicate absorption feature is not properly modeled in these two libraries.}
\label{fig:IRS_gal}
\end{figure}

In Figure~\ref{fig:IRS_gal}, we plot the PAH fluxes of the mock IRS spectra (including the AGN component)
versus those of the observed IRS data. It is clear that, when using the R09 library, the mock PAH fluxes
are generally consistent with observations \citep[also see][]{Lyu2017}, while both the DH02 and DH14
libraries tend to over-estimate PAH fluxes. If the PAH emission is not significantly suppressed in AGN hosts
and can be used as a good star-formation tracer \citep[see, e.g.,][]{Shi2007, Lutz2008, Watabe2008,
Rawlings2013, Esquej2014, Symeonidis2016}, then our results suggest that the R09 library is more 
appropriate to PG quasars than the DH02 or DH14 library. On the other hand, if AGN feedback 
destroys PAH \citep[see, e.g.,][]{A&R1985, Voit1992}, it may not be valid to infer the stellar 
contribution from the PAH emission. Our SED-decomposition results are largely independent of 
the choice of galaxy library (see Figure~\ref{fig:three_median}). Thus our AGN IR SED is more robust than 
the PAH-based results.

\section{Conclusions}
\label{sect:conclusions}
In this work, we revisit the AGN intrinsic IR SED by adopting the SED-decomposition technique. We 
use the AGN template library of \cite{Siebenmorgen2015} as this library contains previous 
commonly-used AGN SEDs \citep[e.g.,][]{Mullaney2011, Symeonidis2016, Lyu2017}. We test 
three galaxy template libraries, i.e., DH02, DH14, and R09. Compared to S16 ($47$ $z<0.18$ 
PG quasars), our final sample (consisting of $42$ sources) contains PG quasars extending to higher 
redshifts ($z<0.5$); in addition, unlike the PAH-based method of S16,
our intrinsic AGN IR SED proves to be insensitive to the choice of galaxy template library.
Furthermore, our fitting results do not have cases where the galaxy component exceeds the observed
total SED as seen in S16. 
Our derived AGN IR SEDs are available in Table~\ref{tbl-2}. The main results of this work are as follows:
\begin{itemize}

\item Through SED decomposition, we derive a median AGN IR SED between 6 $\rm{\mum}$ to 500
$\rm{\mum}$. Our median IR SED is generally consistent with those of M11 and L17, but in contrast
to that of S16 (i.e., S16 underestimated the galaxy FIR contribution, and obtained a very cool AGN FIR SED; 
see Figure~\ref{fig:final}). We speculate that the stellar contribution was not 
well determined/subtracted in S16 because they adopted the DH02 galaxy template library and 
PAH fluxes to determine the stellar contribution. 

\item We find that the AGN IR SED tends to be cooler (i.e., a higher fraction of FIR emission) with
increasing AGN luminosity (Figure~\ref{fig:Lum_dark}). This luminosity-dependent SED 
evolution might be explained if more luminous AGNs tend to have stronger radiative feedback to change 
torus structures and/or their tori have higher metallicities (see Section~\ref{sect:SED_lum}). 

\item Our results do not depend upon the choice of galaxy template library (see
Figure~\ref{fig:three_median}). Meanwhile, we find that the conversion of the PAH emission to
galaxy SED varies with the choice of galaxy template library (see Figure~\ref{fig:pah_three_comparison}).
If the R09 galaxy template library is more appropriate to PG quasar hosts, the PAH fluxes 
predicted by our best fit models are consistent with those observed (Figure~\ref{fig:IRS_gal}). 
\end{itemize}

\section{ACKNOWLEDGEMENTS}
We thank the anonymous referees for their helpful comments that significantly improved the paper.
We thank Yong Shi for providing his PAH fitting code. 
J.X., M.Y.S., and Y.Q.X. acknowledge the support from the China Postdoctoral Science Foundation 
(2016M600485), NSFC-11603022, NSFC-11890693, NSFC-11421303, the CAS Frontier Science Key 
Research Program (QYZDJ-SSW-SLH006), and the K.C. Wong Education Foundation.

\appendix
\section{Energy budget}
\label{sect:energy}
\begin{figure}
\epsscale{0.75}
\plotone{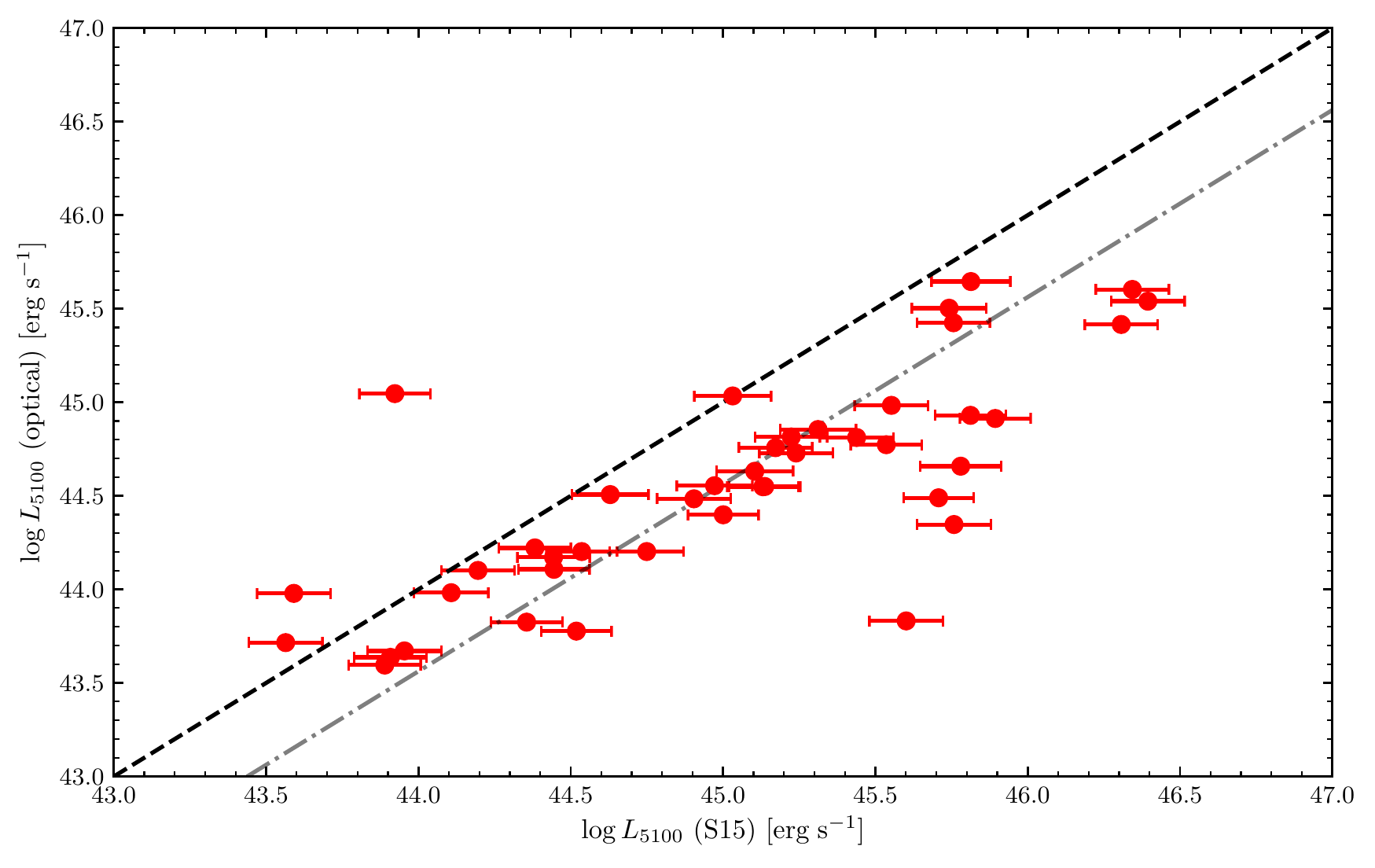}
\caption{Comparison between the optical-derived $5100\ \mathrm{\AA}$ luminosity 
and the IR-derived one. The dashed line represents the one-to-one relation. Overall, the former is smaller 
than the latter by a factor of $2.74$ (indicated by the dash-dotted line), which might indicate the disk 
emission is anisotropic. }
\label{fig:l51-comp}
\end{figure}

In Section~\ref{sect:data}, we interpolate between the SDSS $ugriz$ bands or 
between the Palomar B-band and 2MASS J-band (in the cases of no SDSS counterparts) 
to derive the rest-frame $5100\ \mathrm{\AA}$ luminosity (hereafter $L_{5100}(\rm optical)$). 
The same quantity can also be inferred from our IR decomposition results. First, we use the 
methodology in Section 3.2 of \cite{Siebenmorgen2015} and our best-fitting AGN templates 
to drive the bolometric luminosities for our AGNs (hereafter $L_\mathrm{bol}(\rm S15)$). Second, 
we use the spectral shape of the primary AGN emission adopted by \citet[see their 
Eq.~2]{Siebenmorgen2015} to calculate the rest-frame $5100\ \mathrm{\AA}$ luminosity 
(hereafter $L_{5100}(\rm S15)$). A 
comparison between the two $5100\ \mathrm{\AA}$ luminosities is presented in 
Figure~\ref{fig:l51-comp}. 

There is an evident correlation between $L_{5100}(\rm optical)$ and $L_{5100}(\rm S15)$. 
However, $L_{5100}(\rm optical)$ is systematically smaller (by a median factor of $\sim 2.74$) than 
$L_{5100}(\rm S15)$. This discrepancy might be understood as 
follows. The optical emission 
is widely believed to be produced by the optically thick but geometrically thin accretion disk \citep{SSD}. 
That is, the optical emission is expected to be highly anisotropic and anti-correlate with the 
viewing angle ($\theta$). The line-of-sight effect has been considered in $L_{5100}(\rm S15)$. 
Therefore, the ratio of $L_{5100}(\rm optical)$ to $L_{5100}(\rm S15)$ actually measures 
our viewing angle, i.e., $\theta \sim 68$ degrees. This estimation of the viewing angle is roughly 
consistent with our SED decomposition results (see panel (e) of Figure~\ref{fig:three_parameter}). 
In addition, optical extinction (which effectively reduces our estimations of $L_{5100}(\rm optical)$) 
due to line-of-sight dust might also be partially responsible for the discrepancy. 

\section{Fitting the SEDs with AGN Templates Alone}
\label{sect:agn-alone}
\begin{figure}
\epsscale{0.75}
\plotone{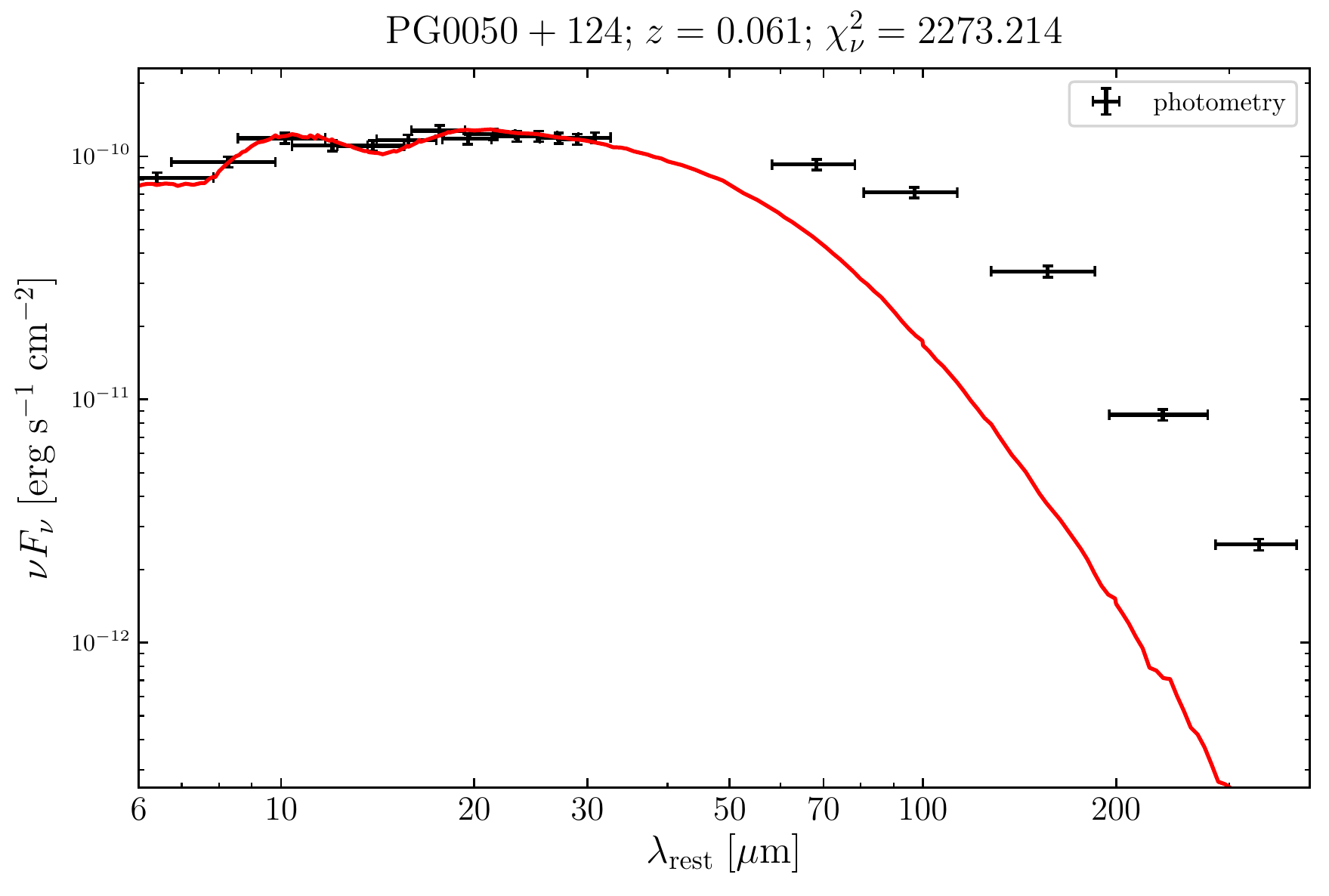}
\caption{An illustration of our SED-fitting results with AGN templates alone. Without a galaxy 
component, the best-fitting AGN template significantly underestimates the observed FIR emission; 
that is, this fit is unacceptable ($\chi^2_\nu=2273.214$). The complete figure set that contains the 
best-fitting results of all 42 sources is available in the online journal.} 
\label{fig:agn-alone}
\end{figure}

For each of the $42$ sources, we also try to fit the observed SED with the AGN templates 
alone (i.e., fixing $c_2\equiv 0$), with one example shown in Figure~\ref{fig:agn-alone}.  
\figsetstart
\figsetnum{2}
\figsettitle{displays the best-fitting results with the AGN templates alone for all our $42$ sources.}
\figsetgrpstart
\figsetgrpnote{The best-fitting results with the AGN templates alone for all our $42$ sources. }
\figsetgrpend
\figsetend
It is clear 
that, without a galaxy component, the best-fitting result is poor and unacceptable ($\chi^2_\nu=2273.214$). 
This is because the best-fitting AGN template cannot account for the FIR data. In fact, for 
each of the $42$ sources, the best-fitting result can be significantly improved if we add a 
galaxy component (see the second and third columns of Table~\ref{tbl-3}). 

\end{document}